\documentclass[fleqn,usenatbib]{mnras}

\usepackage[T1]{fontenc}

\DeclareRobustCommand{\VAN}[3]{#2}
\let\VANthebibliography\thebibliography
\def\thebibliography{\DeclareRobustCommand{\VAN}[3]{##3}\VANthebibliography}

\usepackage{graphicx}	
\usepackage{amsmath}	
\usepackage{amssymb}	

\usepackage{newtxtext,newtxmath}

\usepackage[normalem]{ ulem }
\usepackage{soul}

\newcommand{\frb}{FRB 180916.J0158+65 }

\urldef\hildaurl\url{https://minorplanetcenter.net/db_search/show_by_orbit_type?utf8=%E2%9C%9&orbit_type=8}
\urldef\trojanurl\url{https://minorplanetcenter.net/db_search/show_by_orbit_type?utf8=%E2%9C%93&orbit_type=9}		
\urldef\wikifigureurl\url{https://fr.wikipedia.org/wiki/Groupe_de_Hilda}

\title[Orbiting asteroids for \frb]{Periodic activity from fast radio burst FRB180916 explained in the frame of the orbiting asteroid model}

\author[G. Voisin et al.]{
	Guillaume Voisin$^{1,2}$\thanks{E-mail: guillaume.voisin@obspm.fr;\newline astro.guillaume.voisin@gmail.com},
	Fabrice Mottez$^{1}$,
	Philippe Zarka$^{3}$,
	\\
	$^1$ LUTh, Observatoire de Paris, PSL Research University, CNRS, Université de Paris, Sorbonne Université, 5 place Jules Janssen, 92190 Meudon, France\\
	$^{2}$ Jodrell Bank Centre for Astrophysics, Department of Physics and Astronomy, The University of Manchester, Manchester M19 9PL, UK\\
	$^{3}$ LESIA \& USN, Observatoire de Paris, CNRS, PSL Research University, SU/UP/UO, 92195 Meudon, France 
}

\date{Accepted XXX. Received YYY; in original form ZZZ}

\pubyear{2015}

\begin{document}
\label{firstpage}
\pagerange{\pageref{firstpage}--\pageref{lastpage}}
\maketitle

\begin{abstract}
Observation of fast radio bursts (FRBs) are rising very quickly with the advent of specialised instruments and surveys, and it has recently been shown that some of them repeat quasi-periodically. In particular, evidence of a $P=16.35$ day period has been reported for FRB 180916.J0158+65. We seek an explanation within the frame of our orbiting asteroid model, whereby FRBs are produced in the plasma wake of asteroids immersed in the wind of a pulsar or a magnetar.
   We used the data reported by the CHIME/FRB collaboration in order to infer the orbital characteristics of asteroid swarms, and performed parametric studies to explore the possible characteristics of the pulsar, its wind, and of the asteroids,  under the constraint that the latter remain dynamically and thermally stable. 
   We found a plausible configuration in which a young pulsar is orbited by a main $\sim 10^{-3}M_\odot$ companion with a period $3P = 49$d, three times longer than the apparent periodicity $P$. Asteroids responsible for FRBs are located in three dynamical swarms near the L3, L4 and L5 Lagrange points, in a 2:3 orbital resonance akin to the Hildas class of asteroids in the Solar system. In addition, asteroids could be present in the Trojan swarms at the L4 and L5 Lagrange points. Together these swarms form a carousel that explains the apparent $P$ periodicity and dispersion. We estimated that the presence of  at least a few thousand asteroids, of size $\sim20$km, is necessary to produce the observed burst rate. We show how radius-to-frequency mapping in the wind and small perturbations by turbulence can suffice to explain downward-drifting sub-pulses, micro-structures, and narrow spectral occupancy.
\end{abstract}




\section{Introduction}

Fast radio bursts (FRBs)  consist in short, typically a few milliseconds, and intense flashes that have so far been observed only in radio bands  \citep{petroff_fast_2019}. One of their most puzzling properties  is the large electron column density that the signal has crossed, the so-called dispersion measure (DM), which is encoded in their dynamic
spectrum (e.g. \citet{lyne_pulsar_2012}). The measured DM is compatible with extragalactic and even cosmological distances. 
If the observed DM could in principle be caused by a dense environment, the extra-galactic origin of 12 FRBs has been confirmed by the identification of their host galaxies \citep[see][and references therein]{bochenek_localized_2021}, and in particular the host of \frb \citep{marcote_repeating_2020}.
In addition, the repetition of \frb and at least 19 others \footnote{Twenty repeaters were identified by CHIME/FRB by June 2021: \url{https://www.chime-frb.ca/repeaters}} rules out, in those cases, theories appealing to cataclysmic events such as mergers or collisions. Fast radio bursts caused by Alv\'en wings of planets orbiting pulsars \citep{mottez_radio_2014} have been, to our knowledge, the first theory predicting a periodic repetition of FRBs. Although such perfect periodicity has not been observed so far, we believe that this theory has a few characteristics that are appealing for any theory of FRBs. Alfv\'en wings are akin to a plasma wake left by an electrically conducting object immersed in a magnetised wind. They have been originally theorised and observed in the context of the Jupiter-Io interaction \citep{neubauer_1980}. These wings are favourable sites for radio-emitting plasma instabilities such as the cyclotron maser instability \citep{mottez_radio_2014}. In the particular case of a pulsar the wind is ultra-relativistic with the immediate consequence that any radiation in the wing is highly collimated within a cone of aperture $\sim 1/\gamma$ where $\gamma \lesssim 10^6$ is the Lorentz factor of the wind. Thus, a few interesting characteristics follow: i) compared to isotropic emission scenarios that require tremendous amounts of energy, here little is necessary to produce the observed radio flux, ii) no high-energy counterpart is required
which is in agreement with most of the observations, iii) we know that objects orbiting pulsars are common, only is it unlikely to observe a galactic FRB due to their very a narrow beam, and iv) this very narrow beam explains the short burst duration. Concerning ii) and iii),
we note that the FRB-like event recently observed from the Galactic magnetar SGR 1935+2154 \citep{andersen_bright_2020} was coincident with soft gamma-ray emissions. However, the lack of radio bursts during other high-energy events \citep{lin_no_2020} suggests that the two components could result from two different mechanisms.

We have recently shown (\citet{mottez_repeating_2020}, hereafter MZV20) that Alfv\'en wings of small bodies, such as asteroids and planetoids, can be sufficient to generate FRBs while being sufficiently far from the neutron star to survive evaporation due to the intense irradiation by the pulsar and its environment. 
An Alfv\'en wing is the result of the unipolar inductor created by an electrically conducting asteroid immersed in the magnetised wind of the pulsar: this creates a system of currents coming from the wind and going through the asteroid that propagates along the magnetic field lines. In the pulsar wind the magnetic field is expected to be nearly azimuthal \citep[e.g.][]{petri_theory_2016}, but convected radially with a very large Lorentz factor which sends the current structure at a very small angle to the local radial direction. We assume that the current structure is prone to developing plasma instabilities able to radiate a fraction of the wind power intercepted by the asteroid (MZV20). The plasma being highly relativistic, the radiation is highly collimated in the radial direction. A fast radio burst is then produced when the beam crosses the line of sight of the observer, that is when the neutron star, the asteroid and the observer are aligned. 

This opens the possibility of bursts repeating at seemingly random intervals if asteroids come in belts or swarms. In that latter case, one would expect groups of FRB events occurring periodically within a time window corresponding to the range of orbital phases occupied by the swarm. We call this a swarm transit. Within these swarm transits, each asteroid favourably located along the line of sight may contribute several FRBs in short, apparently random, sequences, as a result of erratic motion of the emission beam in the turbulent pulsar wind. In addition, no strict periodicity between individual events would occur due to motion within the swarm and, depending of the density of objects, some transit windows might contain no event at all. This behaviour seems to be precisely what has recently been reported concerning \frb \citep{collaboration_periodic_2020} but also FRB121102 \citep{rajwade_possible_2020, cruces_repeating_2021} although we shall focus on the former in this work. 

From an observational point of view, if there is some evidence that asteroids could produce some of the observed pulsar timing noise \cite{shannon_asteroid_2013}, possibly impact pulsars \citep[][]{brook_evidence_2014}, or that dust can be found in the environment of a magnetar \citep{wang_debris_2006}, it is not known at present how frequent they are. Searches for planets around pulsars showed that planets do not generally form from the supernova fallback disc \citep[][]{kerr_limits_2015}.
Nonetheless, formation of asteroids remains plausible as there are a number of ways a proto-planetary disc could form around a neutron star \citep[e.g.][]{lin_formation_1991,  nakamura_origin_1991, phinney_pulsar_1993, podsiadlowski_planet_1993}. These works were motivated by the discovery of 3 planets around PSR1257+12 using pulsar timing \citep[][]{wolszczan_planetary_1992, wolszczan_discovery_2012}. Asteroids, however, would usually not be detectable by timing or any other means except as timing noise \citep[][]{shannon_asteroid_2013} or radio variability \citep[][]{cordes_rocking_2008}. Another possibility is the capture of asteroids by close encounter with a star possessing an asteroid belt \citep[][]{dai_repeating_2016} or possibly the tidal disruption of a larger body. 

The orbiting asteroid model distinguishes itself among FRB models by its relatively cheap energy budget. Indeed, although very young pulsars or magnetars are required to explain the brightest bursts (MZV20 and below), only the spin-down power producing the relativistic wind is necessary. Models appealing to magnetar flares require the release of large amounts of magnetic energy which can overpower spin-down by several orders of magnitude, independently of whether radio emission occurs within the magnetosphere \citep[e.g.][]{popov_hyperflares_2010, cordes_supergiant_2016, kumar_fast_2017, wadiasingh_repeating_2019, wadiasingh_fast_2020, lu_unified_2020}, or result from a synchrotron maser generated by a shocked relativistic ejecta in the wind \citep[e.g.][]{popov_millisecond_2013, lyubarsky_model_2014, metzger_millisecond_2017, beloborodov_flaring_2017, beloborodov_blast_2020, yuan_plasmoid_2020}. Another category of models resorts to the release of gravitational and kinetic energy of asteroids colliding with a neutron star \citep[e.g.][]{dai_repeating_2016, bagchi_unified_2017, smallwood_investigation_2019, dai_periodic_2020, dai_magnetar-asteroid_2020} or, alternatively, of matter accreted from a white-dwarf companion \citep{gu_neutron_2016, gu_neutron_2020} which again largely, if briefly, overpowers the spin-down of the neutron star. In the orbiting asteroid model, the periodicity of activity windows is due to the specific orbital dynamics of the system. This is also the case for a number of other models:  \citet{decoene_fast_2020} which is also based on the Alfv\'en wing mechanism, \citet{smallwood_investigation_2019, dai_periodic_2020} for colliding asteroids, \citet{gu_neutron_2020} for accreted white-dwarf companions, and \citet{lyutikov_frb_2020} for either magnetospheric or wind magnetar models. In magnetospheric magnetar models, it has also been proposed that free precession of the neutron star \citep{zanazzi_periodic_2020, levin_precessing_2020}, or ultra-long period magnetars \citep[][]{beniamini_periodicity_2020} might be the cause.

While this paper was under review new bursts were published in \citet{pleunis_lofar_2021} \citep[see also][]{pastor-marazuela_chromatic_2020}. This additional data contains a new set of the CHIME/FRB data which we fully include in our analysis, and low frequency observations from LOFAR and uGMRT. The latter two do not benefit from the same regularity of observation windows which makes them difficult to compare with the CHIME/FRB observations. In addition, their very different frequency band prevents any direct comparison of bursts properties. It also appears that low-frequency bursts occur on average with a delay of 3 days compared to bursts in the CHIME/FRB band. Our model, as it stands, cannot explain such an important delay, although we propose some reflections on the matter in the conclusion of this paper. We also note the observations carried out by uGMRT close to the peak of the activity window in the 550-750 MHz band during three consecutive cycles \citep[][]{marthi_detection_2020}. Unless otherwise stated we do not include these bursts in our analysis in order to ensure homogeneity in observing conditions.

The quasi-daily monitoring of the source \frb by CHIME/FRB over 749 days with a daily exposure window 
of $\sim 15$min has resulted in the detection of 45 events which appear to be bunched in 5-day windows around a period $P=16.35\pm0.18$ days. In this paper, we propose a theory based on \citet{mottez_radio_2014} and MZV20 compatible with the reported observations of FRB 180916.J0158+65. We also propose a set of predictions that will be falsifiable in the near future provided regular observations of this FRB are continued.  We suggest mechanisms to explain the characteristics of the bursts.

\section{Trojans and Hilda-type asteroid swarms}
\begin{figure}
\centering
\includegraphics[width=0.5\textwidth]{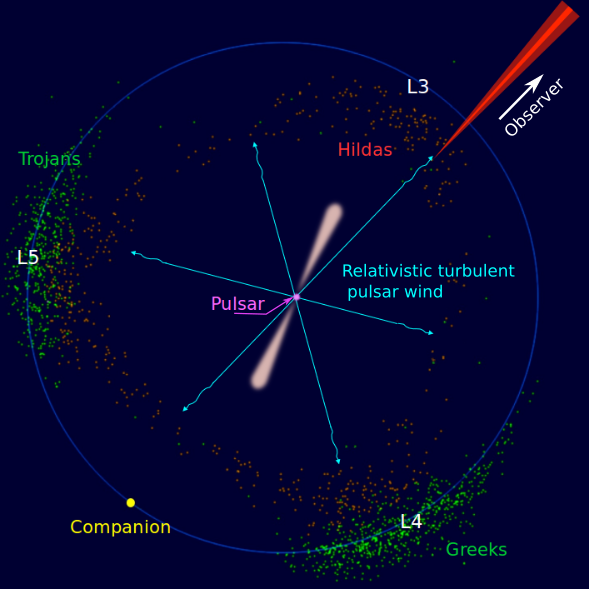}
\caption{\label{fig:sketch} Sketch showing the location of the Trojan/Greek (in green) and Hilda (in orange) asteroid swarms relative to the pulsar (at the centre) and its main companion (in yellow). In the Solar system, the latter would be Jupiter and the former the Sun. The pulsar wind (cyan arrows emerging from the pulsar) interacts with asteroids to form Alfv\'en wings: plasma wakes where plasma instabilities producing radio emissions can develop. They are highly collimated ($\sim 1''$) due to the high wind Lorentz factor, and only a single radio beam can plausibly cross the line of sight of an observer at a given time, thus creating a fast radio burst. The direction of the beam wanders randomly within a wider angle (red shaded area) due to turbulence in the wind (wiggles on wind arrows) which can result in multiple bursts within short time scales. Adapted from Fig. 1 in \wikifigureurl.}
\end{figure}

\subsection{Asteroids with period P}
Let us assume that a swarm of asteroids orbiting a pulsar is indeed responsible for the observed bursts of \frb. Then, this swarm is orbiting at $P = 16.33\pm0.12$ days, which is the orbital period favoured by the search realised in \citet{pleunis_lofar_2021}. The 5-day window then implies that the swarm covers more than one quarter of its orbit at a distance $\sim 0.14$AU from the neutron star (assuming a 1.4 solar masses). Such a swarm would be gravitationally unbound as can be seen from the fact that a companion with a Roche lobe covering that much of the orbit would require a body more massive than the neutron star itself. Although one could invoke the disruption of a planet due to tidal forces and/or overheating by the pulsar wind \citep{kotera_asteroids_2016}, the probability of catching such an event seems unlikely. An asteroid arc following a disruption would diffuse until it forms an asteroid belt unless some very special dynamical mechanism keeps it stable. Let us note that, although unlikely, such arcs -- named Libert\'e - Egalit\'e - Fraternit\'e -- exist around Neptune. They are highly dynamic, and maybe unstable \citep{Sicardy_1992,dePater_2005}.

\subsection{Trojans with period 3P}
\label{sec:trojans}
In the Solar system the Trojans asteroids are co-rotating with the L4 and L5 Lagrange points of the Sun-Jupiter system (the swarm around L4 is also called the Greek camp by opposition to the Trojan camp at L5). Each swarm spans $\sim90^\circ$ in orbital phase \citep{levison_dynamical_1997} and more than $20^\circ$ in inclination \footnote{See the Trojan page of the Minor planet Center: \trojanurl}. In that respect, Trojan swarms accompanying what we shall call the main pulsar companion (see Fig. \ref{fig:sketch}) are good candidates for explaining FRBs within our theory but, unless the two swarms are highly asymmetric, two transit windows separated by $120^\circ$ should be seen. However, given the small number of events, we simulated that two Trojan swarms with an orbital period of $3P$ could mimic an apparent period of $P$ as it would not be very different from an orbit with three equidistant swarms where the third one has been missed by (lack of) chance. Nonetheless, folding the \frb events at $3P$ should then show only two groups, and not three as it happens (see Fig. \ref{fig:frbprop}). 

The stability of asteroids at L4 and L5 sets a constraint on the mass ratio $m_c / m_p \lesssim \zeta_T \equiv 0.04$ (see e.g. \citet{beutler_methods_2004} eq. 4.142), where $m_c$ is the mass of the companion and $m_p$ the mass of the pulsar. For a typical $m_p \sim 1.4M_\odot$ this gives the upper bound $m_c \lesssim 0.056 M_\odot$. This corresponds to substellar objects such as a brown dwarf, an ultra low-mass white dwarf, black widow companions, or a planet.  

\subsection{Hildas with period 2P}
There exists another class of asteroids in the Solar system called the Hildas which does have the property of forming three equidistant swarms just inside the L3, L4 and L5 points \citep{broz_asteroid_2008}. These asteroids share an orbital period around the Sun of approximately $2/3$ of that of Jupiter and have moderately eccentric orbits with $e\lesssim 0.3$ \footnote{See Hildas at Minor Planet Centre: \hildaurl}. Hence, they undergo a 3:2 resonance with Jupiter such that their aphelia is successively near each of the three Lagrange points over three orbital periods. Contrary to the Trojans, they do not follow the Lagrange points but their stream accumulates near these points thus creating apparent swarms. 
Three identical equidistant swarms (effectively) orbiting at $3P$ would be virtually impossible to distinguish from a single entity at $P$, unless the main companion's mass is sufficiently large such that the L3 point be significantly closer to the pulsar than the two others, in which case the wind magnetic field would be larger at that point and create more intense, and therefore possibly more numerous, observable bursts. Another possibility, which we favour hereafter, is that both Hildas and Trojans co-exist in the system, thus making the population of asteroids denser along the line of sight of the observer at L4 and L5 than at L3 as depicted in Fig. \ref{fig:sketch}. 

In fact, it would seem rather arbitrary that only one type of asteroid exists and therefore more likely that both are present, if any. In the Solar system, these asteroids lie just outside the main belt in terms of their semi-major axis.  We will see below why this is not a configuration occurring by chance. Note that bursts might also be created by the main companion, only is it very unlikely that our line of sight crosses its Alfv\'en wings in particular. On the other hand, if its orbital orientation was favourable one would see this particular burst repeating with a more accurate periodicity \citep[][]{mottez_radio_2014}.

\subsection{Stability of asteroid configurations and companion mass
\label{sec:orbstab}}

\begin{figure}
	\centering
	\includegraphics[width=\columnwidth]{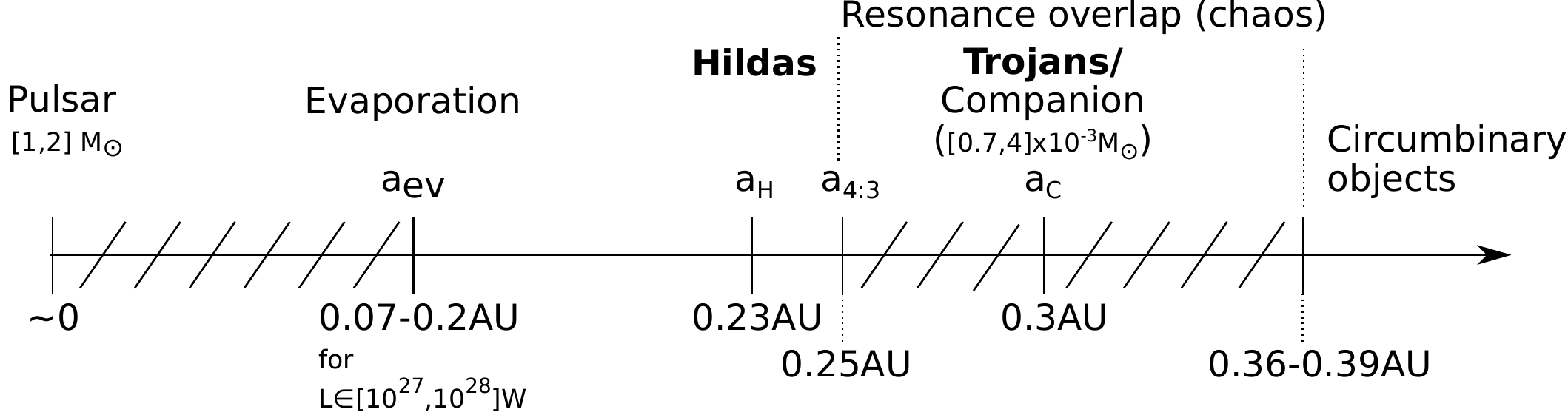}
	\caption{Summary of section \ref{sec:orbstab}. The horizontal axis shows the semi-major axis of each body (and not the distance to the pulsar which might vary because of eccentricity). Hashed areas represent unstable ranges. }
	\label{fig:orbconf}
\end{figure}
The periodicity of the observed bursts imposes the presence of Hilda asteroids with possibly the presence of Trojans, but also the absence of a substantial asteroid belt which would otherwise produce homogeneously distributed bursts. The possibility of various configurations depends essentially on two parameters : the mass ratio $\zeta$ and the pulsar high-energy luminosity $L$. Note that the power emitted by the pulsar under the form of an electromagnetic wave at the spin period, as per a rotating dipole, is not included in $L$ because its wavelength being much larger than an asteroid it does not interact with it \citep{kotera_asteroids_2016}. It follows that $L$ is less than the spin-down power of the pulsar, although gamma-ray observations have shown that it can be a significant fraction of it \citep[e.g.][]{collaboration_second_2013, guillemot_gamma-ray_2016}. 

We consider that an asteroid survives if it can thermally re-radiate the energy it absorbs from the pulsar without melting, namely if
\begin{equation}
    L\frac{\pi R_a^2 }{4\pi r^2} < 4 \pi R_a^2 \sigma T_{\max}^4,
\end{equation}
where, as in \citet{mottez_repeating_2020}, we take $T_{\max} \sim 1400$K the fusion temperature of iron, $R_a$ is the asteroid radius, $r$ the distance t the pulsar, and $\sigma$ is Stefan-Boltzmann's constant. 
Taking $r = a_H (1-e_H)$ with $a_H = a_c (2/3)^{2/3}$ the Hildas' semi-major axis, $a_c$ the companion's, and $e_H$ the orbital eccentricity of Hildas asteroids, we get 
\begin{equation}
    L < L_{\max} \equiv 16 \pi \sigma T_{\max}^4(2/3)^{4/3} a_c^2 (1-e_H)^2 \simeq  1\times 10^{28} \mathrm{W} (1-e_H)^2,
\end{equation}
assuming a pulsar of $1.4M_\odot$ and an orbital period of $3P$. Hildas orbits are (gravitationally) stable up to $e_H \sim 0.3$ for which $L_{\max} \simeq 6\times 10^{27}$W is required. For $L_{\max}(e_H=0.3) = 6\times 10^{27}\mathrm{W} < L < L_{\max}(e_H=0) = 1\times 10^{28}$W Hildas asteroids are still possible but their eccentricity distribution is truncated by evaporation.

Given the luminosity $L$, we may define $a_{\rm ev}$ the distance of closest approach without evaporation,
\begin{equation}
    a_{\rm ev} = \sqrt{\frac{L}{16 \pi \sigma T_{\max}^4}} \simeq 0.07 {\rm AU} \left(\frac{L}{10^{27}\mathrm{W}}\right)^{1/2}.
\end{equation}
In principle asteroids with shorter periods than Hildas's may exist in the interval $[a_{\rm ev}, a_H]$ provided $L < L_{\max}(e_H = 0)$. These would be responsible for bursts homogeneously distributed across orbital phase, and presumably stronger because of the proximity of the pulsar. The absence of detection of such bursts can be explained by a nearly maximal $L$, which seems plausible since most of the values compatible with FRB production lie in the range $[10^{27}, 10^{28}]$W (see below and appendix \ref{etude_parametrique}). Alternatively, one has to consider a formation mechanism favouring the production of asteroids with orbital periods close to that of the companion, such as perhaps the debris of a collision between the companion and another large object.

The Hilda family of asteroids enjoys a particular stability while most other low-order resonances in the main belt are subject to chaos and instability leading in particular to the Kirkwood gaps \citep[e.g.][]{moons_review_1996, michtchenko_comparative_1996}. Note that the fundamental behaviour of resonances in the restricted three-body problem depends on the semi-major-axis ratio of and the mass ratio \citep{murray_solar_1999} and therefore can be extrapolated here. In particular, the absence of asteroids between the Hildas's orbits and the companion suggests that, similarly to the Solar system, chaotic motion dominates in that region. This particularly happens when first-order mean-motion resonances overlap. In the following, we assume small mass ratio, eccentricity and inclination of the companion, which allows us to use simple analytical estimates \citep{petit_amd-stability_2017} of the semi-major axis $a$ for which resonance overlap occurs,
\begin{equation}
\label{eq:overlap1}
    1 - \frac{a}{a_c} < C \zeta^{2/7},
\end{equation}
where $C$ is a numerical constant in the interval $[1.3, 1.5]$ \citep{petit_amd-stability_2017}. 
This, in turn, implies that Hilda asteroids can only exist if the mass ratio is sufficiently small to allow for a stable 3:2 resonance, that is if
\begin{equation}
    \zeta < \zeta_H \equiv \left(C^{-1} (1 - \frac{a}{a_c})\right)^{7/2} \simeq 2\times 10^{-3}  \left(\frac{C}{1.4}\right)^{7/2}.
\end{equation} 
In order to prevent any stable orbit in the region $[a_H, a_c]$ the mass ratio should also be sufficiently large so that overlap starts around the next first order resonance (as in the Solar system in fact), that is $3:4$ for which $a_{3:4}/a_c = (3/4)^{2/3}$, leading to 
\begin{equation}
    \zeta > \zeta_{\min} \equiv \left(C^{-1} (1 - \frac{a_{3:4}}{a_c})\right)^{7/2} \simeq 7\times 10^{-4} \left(\frac{C}{1.4}\right)^{7/2}.
\end{equation} 
Let us remark that $\zeta_H < \zeta_T$ and therefore if Hildas are in a stable region, that is if $\zeta<\zeta_H$, then $\zeta < \zeta_T$ and Trojans are stable as well (see Sec. \ref{sec:trojans}). Thus in the following both families will be considered. Furthermore taking a pulsar mass in the range $[1,2]M_\odot$, and considering $\zeta_{\min} < \zeta < \zeta_H $, we can constrain the companion mass to the range 
\begin{equation}
    7\times 10^{-4} M_\odot \left(\frac{C}{1.4}\right)^{7/2} < m_c <  4\times 10^{-3} M_\odot \left(\frac{C}{1.4}\right)^{7/2}.
\end{equation}

Circumbinary asteroids are similarly excluded of a chaotic zone beyond the companion's orbit. For $a>a_c$, Eq. \eqref{eq:overlap1} becomes \citep{petit_amd-stability_2017}
\begin{equation}
        1 - \frac{a_c}{a} < C \zeta^{2/7}.
\end{equation}
Orbits are not subject to resonance overlap provided that 
\begin{equation}
    a > a_{\rm cb} \equiv a_c \left(1 - C \zeta^{2/7}\right)^{-1},
\end{equation}
where $a_{\rm cb}$ is the lower limit for stable circumbinary objects. Using the previously obtained contrains on the mass ratio we obtain $a_{\rm cb}/a_c \in [1.2, 1.3]$ fo $\zeta \in [r_{\min},\zeta_H]$, or $a_{\rm cb} \in [0.36,0.39]$AU (taking $C=1.4$). 
The FRB flux density decreases as an inverse square law of the distance to the pulsar \citep{mottez_repeating_2020}, assuming a constant wind Lorentz factor and identical asteroids. If the flux density at $a_H$ is $S_H$, it drops down to $S_c \simeq 0.6S_H$ at $a_c$ and $S_{\rm cb} \in [0.34, 0.40]S_H$. $S_c$ gives an estimate of the flux radiated by a Trojan or a Hilda asteroid with large eccentricity. $S_{\rm cb}$ is an upper limit for the flux of potential circumbinary asteroids. Thus, even if circumbinary objets exist in the system they might not be detected due to a lower flux.

\section{Burst statistics\label{sec:burststat}}

Following the previous section, we assume for the remainder of this article that the 45 FRBs reported by CHIME/FRB \citep{collaboration_periodic_2020, pleunis_lofar_2021} originate from three swarms of asteroids located at the L3, L4 and L5 Lagrange points, as depicted in Fig. \ref{fig:sketch}, of a companion orbiting a magnetized neutron star with a period $3P=49$ days. We associate to each swarm a transit window of $\sim 5$ days which has been observed in five daily 15min exposures (see also Fig. \ref{fig:burstLpoint}). As stated in the introduction, we do not include here the low-frequency bursts observed by LOFAR and uGMRT \citep{pleunis_lofar_2021} nor the uGMRT 550-750 MHz observations \citep[][]{marthi_detection_2020}, in order to take advantage of the homogeneity of the observing conditions and setup.

\subsection{Burst multiplicity and identification of asteroid transits}
\label{sec:asteroidid}
We note that bursts are usually not seen across all five daily
exposure windows corresponding to a swarm transit. We call ``active exposure windows'' those exposures during which at least one burst was recorded. Although in principle each individual burst could be due to a different asteroid, we have shown in MZV20 that it is possible that each asteroid results in several bursts bunched in a ``wandering'' time interval $\tau_w$ of order $1$h. This is due to the wandering motion of the source in the turbulent pulsar wind which results in the narrow beam sweeping a much wider area and possibly crossing several times the observer's line of sight. The extent of the area covered by the wandering beam translates into a characteristic wandering time during which the beam is susceptible to cross the line of sight and bursts detections remain possible. In principle, one can infer $\tau_w$ from the bunching of observed bursts. Here, the small number of bursts together with the short, 15 min, contiguous observations makes any determination highly uncertain. However, we note that bunches of up to four bursts were observed within a single 15 min exposure window, and that in eleven out of the twenty non-empty transit windows the totality of the bursts for each of these windows occurred within a single exposure (out of five per transit window), see Fig. \ref{fig:burstLpoint}. In addition, turbulence is expected to be only a small perturbation to the bulk radial motion of the wind, which implies that $\tau_w \ll 3P$. For these reasons, we can assume that $1\mathrm{day} > \tau_{\rm w} \gtrsim 15 \mathrm{min}$ and therefore that all the bursts occurring during a single 15 min exposure result from the transit of a single asteroid. 

Within these conditions, we can check that the source wandering motion is only a small perturbation of the bulk radial motion. As a plausible value, let us take $\tau_w \sim 1$h. The wandering of the beam by an angle $\alpha_w$ results i) from an actual transverse displacement of the source within a radially flowing wind, and ii) from the changes in the direction of the velocity of the radiating plasma blob, which directly translates into emission direction due to relativistic beaming. In i), the condition for the beam to keep crossing the observer's line of sight during $\tau_w$ is that $\alpha_w$ be larger than the angle travelled by the asteroid on its orbit, that is $\alpha_w \gtrsim n_{\rm orb} \tau_w \sim 2\times 10^{-3}$rad, where $n_{\rm orb} \equiv 2\pi/P_{\rm orb} = 2\pi/3P$. In this case, the transverse displacement velocity of the source is $v_\perp \gtrsim a n_{\rm orb} \sim 2\times 10^{-4}c$, which is only a small perturbation to the bulk velocity of the wind given by $v_\parallel\simeq c$. In case ii), $\alpha_w = v_\perp/v_\parallel$. Similarly to i), bursts remain visible during $\tau_w$ if $\alpha_w \gtrsim n_{\rm orb} \tau_w$ which leads to a larger minimum transverse velocity, $v_\perp \gtrsim 2\times 10^{-3}c$, but still at perturbation level. Unfortunately, theoretical constraints on the level of turbulence and inhomogeneity that far in the wind are lacking, partly due to the computational challenge it represents. \citet{cerutti_dissipation_2017} carried out calculations as far as a hundred light-cylinder radii by focusing on the current sheet region\footnote{The current sheet is the region of magnetic inversion generated by the magnetic equator of the pulsar, and responsible for the stripped structure of the wind as it oscillates across the spin equator \citep{petri_theory_2016}}. The results show much larger levels of fluctuations than necessary for the lower limits of the present discussion (see in particular their Fig. 10, where the random deflection of test particles relative to the radial direction is so large that it is visible by eye). Asteroids are located between $10^3$ and $10^5$ light-cylinder radii, depending mostly on the pulsar spin period, meaning that turbulence may have enough time to develop.
In total 33 asteroid transits were observed by CHIME/FRB (see  Fig. \ref{fig:burstLpoint}). We also note that turbulence naturally explains the two pairs of bursts separated by only 60ms that were observed and which were thus
connected to the same object and not independent. 

\subsection{Identification of asteroid swarms \label{sec:idast}}
It readily appears that more than half of the swarm transit windows do not show any event (see Fig. \ref{fig:burstLpoint}). This may simply mean that no asteroid was transiting at all, consistent with the fact that often a single asteroid was seen during a transit window. 
However, it is also possible that an independent mechanism is making bursts visible only momentarily. As pointed out in \citet{spitler_2018} for FRB121102, a possible mechanism is Galactic scintillation. In most cases this causes an attenuation of the intrinsic flux of the source and more rarely enhances it (see e.g. \citet{cordes_fast_2019}). However, given that the source has a low Galactic latitude of $3.73 \deg$, its Galactic scintillation bandwidth is limited to at most a few kHz at the central observing frequency of 600 MHz according to the NE2001 model \citep{cordes_ne2001i_2002}. This is several orders of magnitude smaller than the observing bandwidth (400 MHz) and therefore no significant modulation is expected due to Galactic scintillation. It is not clear whether another hiding mechanism could play a role, but we stress that although we cannot dismiss this possibility, it is not a necessary part of our model.

\begin{figure}
	\centering
	\includegraphics[width=\columnwidth]{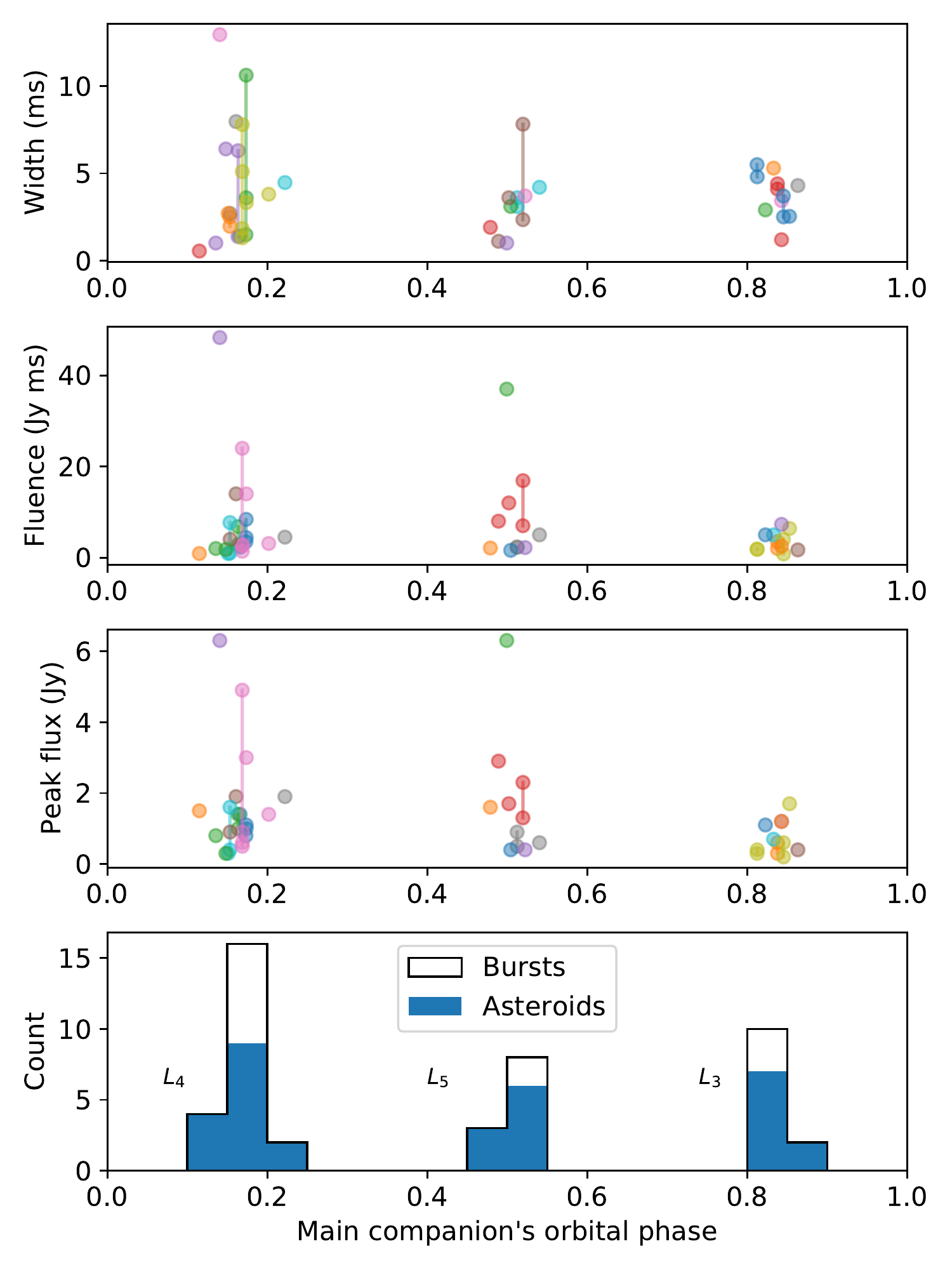}
	\caption{Summary of the properties of the bursts of \frb observed by CHIME/FRB as reported in \citet{collaboration_periodic_2020} and \citet{pleunis_lofar_2021} as a function of the main companion's orbital phase, assuming an orbital period of $3P = 49$ days. Bursts associated to a single asteroid (and coincidentally to a single exposure window) have the same colour and are connected to each other by a solid line. The bottom panel shows the burst count (black line) and the number of asteroids (blue bars) vs the orbital phase of the main companion.} The most probable Lagrange point associated with each group of bursts is indicated on the bottom panel.
	\label{fig:frbprop}
\end{figure}

\begin{figure}
	\centering
	\includegraphics[width=\columnwidth]{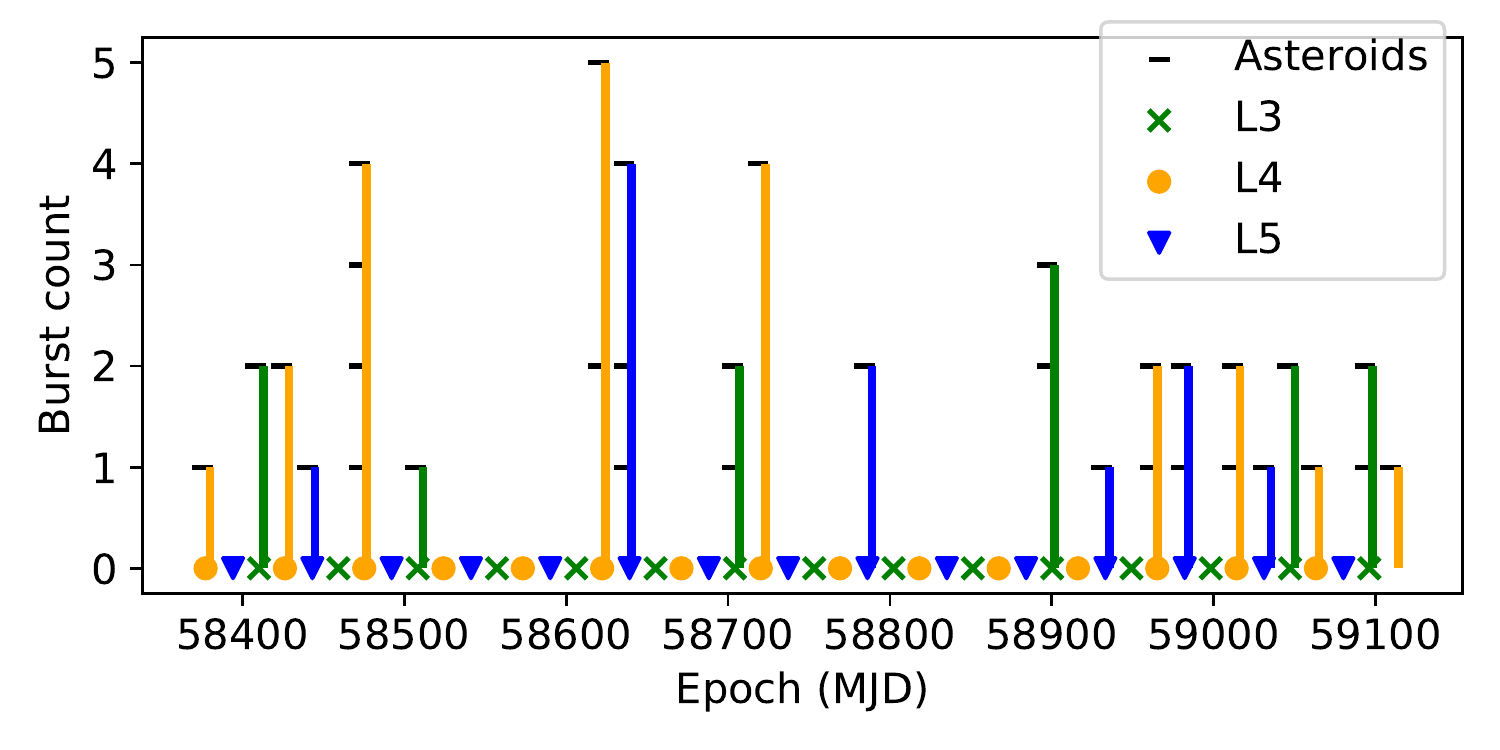}
	\caption{Number of bursts seen during each swarm transit window in the time span of the CHIME/FRB observation reported in \citet{collaboration_periodic_2020, pleunis_lofar_2021}. Each transit window lasts 5 days and was observed for $\sim1$h$15$min ($\sim 15$min/day exposure) totalling 13h45min of exposure during transits on the entire observation span. Each black tick corresponds to one ``active'' exposure window translating into as many asteroids if one assumes a beam wandering time $15\mathrm{min} \lesssim \tau_{\rm w} < 1\mathrm{d}$}

	\label{fig:burstLpoint}
\end{figure}

Although the number of asteroid transits seen at each Lagrange point might be the result of a random fluctuation, or of some hiding mechanism, we assume for convenience in Fig. \ref{fig:frbprop} and \ref{fig:burstLpoint} that L3 corresponds to the smallest number of detected
transits, 11, as well as the smallest dispersion in peak flux, fluence and width, since it is the point where only Hilda asteroids can be seen. It follows that L4 contains the largest number of transits with a total of 14, and L5 had 11 (but with somewhat more dispersed characteristics than L3). We also note that L3 has the narrowest swarm transit duration as the bursts are distributed over 2 day, compared to 3 and 5 days for L4 and L5 respectively. This is somewhat correlated with the number of asteroids within each swarm, but might also be indicative of their intrinsic sizes. 

We note that the observations carried out by uGMRT during 2h within three consecutive activity windows between 550 and 750 MHz \citep[][]{marthi_detection_2020}, a sub-band of CHIME/FRB, revealed 0, 12 and 3 bursts at our conventional L4, L5 and L3 points respectively. If included, this would translate in respectively 0, 12 and 2 transits (2 bursts were separated by less than 100 ms, see above), which would not change the order derived from CHIME/FRB data and even strengthen it by balancing L5 with L4.

In this model, the properties of successive bursts, although generated by the same asteroid, should vary randomly from one burst to the next. The reason for this is that the intersection of the beam with the line of sight varies randomly. In order to verify that, we computed the normalised cross-correlation $c_x = \left<\Delta x_i \Delta x_{i+1}\right>/\left<x_i\right>\left<x_{i+1}\right>$, where $ \Delta x_i = x_i - \left<x\right>$ is the difference with the mean of property $x$ of burst $i$ in a given ``active'' exposure, is small. The
property can be either its fluence $F$, peak intensity $I$, or width $W$. In order to account for measurement errors, we drew 10,000 samples assuming each property to be Gaussian distributed with mean and standard deviation given by its measured value and error. We got $c_F = 0.07 \pm 0.07, c_I = 0.07 \pm 0.08, c_W = -0.2 \pm 0.03$, where error bars delimit the 68\% confidence region. Correlations are consistent with zero for fluence and peak intensity, and weak but significant for width. These weak or null correlations are consistent with our hypothesis. On the other hand, significant correlations would be expected if these bunches were related to an eruption-like mechanism, where aftershocks of lower intensity usually follow a primary event (see e.g. \citet{aschwanden_25_2016} for a discussion of this type of mechanism in the framework of self-organised criticality). 


\subsection{Periodicity}

We searched for asymmetries between the three Lagrange points that could reveal a $3P$-periodicity. In particular, we performed pairwise two-sample Kolmogorov-Smirnov tests between the burst counts received at each Lagrange point (bottom panel of Fig. \ref{fig:frbprop}), thus trying to assert if these three samples derive from the same distribution. The results depend heavily on the number of exposures that are assumed hidden, but assuming all are active their burst rates are consistent with a single distribution common to all three Lagrange points (p-value larger than 99\%). Similarly, we compared the three samples of width, peak intensity, and fluence (Fig. \ref{fig:frbprop}) and obtained inconclusive p-values ranging from $0.1$ to $0.6$ for the L3/L4 and L3/L5 comparisons, but respectively 0.63, 0.97 and 0.83 for the L4/L5 comparison, thus suggesting a common peak and fluence distribution between L4 and L5. This strengthens our identification of asteroid swarms (Sec. \ref{sec:idast}), since L4 and L5 are expected to be more similar. Mostly, it illustrates the fact that unless there is a striking difference between the three swarms, the small sample sizes prevent any conclusion. In fact, we checked by simulation that even if one swarm was missing altogether, a period of $P$ could still be favoured against a period of $3P$ in a periodogram such as the one reported in \citet{collaboration_periodic_2020}. Indeed, due to the small sample, removing a swarm can be indistinguishable from a statistical fluctuation. Of course, in this case only two groups of bursts would appear on Fig \ref{fig:frbprop}.

\subsection{Number of asteroids}
Estimating the number of asteroids in the system requires further assumptions concerning the properties of the swarms. Following MZV20, we considered a simplistic model where the swarms cover an angle $\alpha =0.1 \mathrm{rad}$ in inclination, akin to the angle covered by the Trojans of the Solar system. 
The effective area covered by the emission beam is dominated by wandering in the turbulent pulsar wind, and we approximated it to a cone of aperture $\alpha_{\rm w} \sim 2\pi \tau_{\rm w}/ P_{\mathrm{orb}}$, and $P_{\rm orb}$ is the orbital period. The visible asteroids are those contained within a band of thickness $\alpha_{\rm w}$ in inclination. Assuming a uniform distribution of asteroids within three identical swarms one gets the total number of asteroids $3N_{\rm swarm} = N_v \alpha/\alpha_{\rm w}$, where $N_{\rm swarm}$ is the number of asteroids in one swarm and $N_{\rm v}$ is the number of visible asteroids during one orbital period. This relates to the average rate of transiting asteroids $n_{\rm a} = N_{\rm v}/P_{\rm orb}$ such that 
\begin{equation}
N_{\rm swarm} = 1.2\times 10^3\left(\frac{n_{\rm a}}{4\mathrm{d}^{-1}}\right)\left(\frac{\alpha}{0.1\mathrm{rd}}\right)\left(\frac{P_{\rm orb}}{49.05\mathrm{d}}\right)^2\left(\frac{\tau_{\rm w}}{1\mathrm{h}}\right)^{-1},
\end{equation}
where we estimated $n_{\rm a}$ from 33 asteroids (see Fig. \ref{fig:burstLpoint}) seen over 15 min of observation during 749 days. This rate could be somewhat higher if one assumed that a number of asteroids were missed due to some hiding mechanism (Sec. \ref{sec:idast}). On the other hand, it is possible that some asteroids have been transiting several times, in which case this rate is overestimated. We have also used $P_{\rm orb} = 3P$ which, although not the orbital period of Hilda asteroids, is the effective period over which the three swarms are transiting. 

For comparison, the number of asteroids larger than 1 km at the L4 Jupiter Trojan swarm is estimated to be $\sim 1.6\times 10^5$ for a total mass of $\sim 10^{-4} M_\mathrm{Earth}$ \citep{jewitt_population_2000}.
About 99\% of these asteroids are smaller than 10 km, while $\sim 10^3$ (1\%) are between 10 and 20km in diameter \citep{jewitt_population_2000}. Therefore, in the following we focused our parametric study on ``small'' asteroids ($R_c \leq 10$km) as the most likely candidates for FRBs. This is because we do not know the actual size distribution.
Although the volume of dynamical stability of the swarms is not a straightforward problem \citep[e.g.][]{levison_dynamical_1997} we note that the orbit of Jupiter is only $\sim 20$ times wider than the orbit assumed here and therefore the (angular) density of asteroids does not need to be larger if the swarms span a similar angular size. This contrasts with the simulations carried out by \citet{smallwood_investigation_2019} in the context of the colliding asteroid model \citep{dai_repeating_2016}, which showed that an asteroid belt with a density several orders of magnitude larger than that of the solar system's main belt or Kuiper's belt was necessary for that model.

\section{Individual burst properties in the orbiting asteroid model}
 \label{sec:individualbursts}
In this section, we recall the main ideas of the FRB mechanism assumed in this paper, and expand beyond MZV20 in order to propose possible explanations to some of the observed characteristics of the bursts from \frb \citep{collaboration_periodic_2020,2020ApJ...896L..41C, pleunis_lofar_2021}. Their peak flux densities range from a few tenth of Janskys to about 10 Jy in all frequency bands. The typical duration of a burst is a few milliseconds, which have occasionally been split into several sub-bursts of similar duration. The peak frequency of these sub-bursts is drifting to lower frequencies as times passes. This characteristic has notably been seen in FRB121102 and dubbed the "sad trombone effect" \citep{hessels_frb_2019} as well as in other repeaters \citep{fonseca_nine_2020}. The burst bandwidth appears to be relatively narrow, typically $\Delta \omega/\omega < 0.5$ within the CHIME/FRB band (400 - 800 MHz) as well as within the LOFAR band (120 - 180 MHz), which is also a property that seems to be shared by other repeaters \citep{hessels_frb_2019,fonseca_nine_2020, kumar_extremely_2020}. On top of the sub-bursts, micro-structure with characteristic timescale of tens of microseconds has been evidenced \citep{nimmo_highly_2020} as well as in other repeaters \citep{farah_frb_2018, cho_spectropolarimetric_2020}. Polarisation appears to be 100\% linear in the CHIME/FRB band, while a depolarisation down to 30\% has been observed in the LOFAR band, which could possibly be due to scattering without ruling out an intrinsic cause \citep{pleunis_lofar_2021}. A significant rotation measure, $\sim -115\rm rad/m^2$ has been observed \citep{collaboration_periodic_2020, 2020ApJ...896L..41C, pleunis_lofar_2021}.

\subsection{Burst flux density: pulsar and asteroid properties}
We conducted a parametric study in order to see if the MZV20 model fits the observed flux densities of FRB 180916.J0158+65 when we set the companion orbital periods to $P$, $2P$ and $3P$. We tested various parameter sets  with the same equations and the same constraints as in MZV20. 
Apart from the orbital periods $P_{\rm orb}$ and the distance $D$ to the observer, estimated at $D=0.15$ Gpc \citep{marcote_repeating_2020}, the choice of the parameter set is the same as in MZV20 where it is discussed. Out of 1,336,500 parameter sets tested for a pulsar, 140,382 of them could provide FRBs above 0.3 Jy, for companions (small or large) that do not evaporate. 
When we restricted our parameter space to asteroids of diameter $R_c < 10$ km, associated with a pulsar emitting more than $10^{27}$ W in the form of non-thermal photons, with a spin-down age $\tau=P_*/\dot P_*$ larger than 10 years, we found 1,251 solutions. Had we chosen a larger maximal radius, the number of solutions would also be more important. Details are given in appendix \ref{etude_parametrique}.
This study shows that \frb is compatible with young magnetized pulsars with short spin periods (3 - 10 ms), of ages $\tau=P/\dot P$ comprised between 10 years and 1000 years, surrounded by metal rich asteroids of size $R_c \le 10$ km having orbital periods $P$, or $2P$ or $3P$. Those asteroids do not evaporate, therefore the duration of such systems as sources of FRBs is mainly constrained by the pulsar spin down, or said differently, by their spin-down age $\tau$. This means that according to the present model, repeating FRBs could be observed from a few decades to a millenium at the present flux level. 

This is at odds with the recent estimate that the age of the source \frb would be $>10,000$ years \citep{tendulkar_60_2021}. However, this estimate relies on the observation that the source lies at the edge of a star-forming region, and assuming the source was born at the centre with a typical neutron-star kick, the authors compute the time needed to reach its current position. In fact, one cannot rule out that the source was born much closer to its current position. The argument can even be reversed: assuming a typical neutron-star kick and that the source was born somewhere within the star-forming region, one may conclude that the fact that it is currently still within that region bounds its age to no more than $\sim 2\times 10,000$ years, the bound being reached if the birthplace is diametrically opposed to the current location of the source.

\subsection{Burst duration}
The radiated power depends on the local magnetic field intensity, the wind Lorentz factor, the size of the asteroid, its conductivity, as well as the intrinsic efficiency factor of the yet unspecified radiation mechanism (MZV20). The burst duration is given by the time needed for the beam to cross the line of sight of the observer, that is the total angular size of the beam $ \alpha$ (as seen from the pulsar) divided by the angular velocity $\dot \varphi$. The angular size is the sum of the size of the source, similar to the size of the asteroid, and of the aperture of the beam due to the combination of intrinsic and relativistic beaming. The angular velocity of the source is due partly to its orbital motion, and mostly to turbulence in the wind which randomly moves the Alfv\'en wing structure about its central position. Thus, for an asteroid of size $2R_a\sim20$km, Lorentz factor $\gamma \sim 10^6$, distance to the pulsar of $r\sim 0.3$AU (see Fig. \ref{fig:orbconf}), and characteristic burst duration $\tau\sim 5$ ms one gets $\dot \varphi \sim 10^{-4}$rad/s corresponding to a transverse (turbulent) velocity of $v_\perp \sim 0.015c$ (MZV20). Note that this is an upper limit, since one does not expect the source to cross the line of sight through its widest cross-section, and it is also consistent with the constraints required in Sec. \ref{sec:asteroidid} to explain multiple bursts by multiple crossings of the same beam.

\subsection{Sub-bursts, bandwidth, downward frequency drifting, and micro-structure \label{sec:dfd}}
In the previous section, we considered the global motion of the Alfv\'en wing structure driven by orbital motion and large scale turbulence. In this section we argue that the detailed structure of the bursts should be seen as a consequence of smaller-scale turbulence. 

\begin{figure}
    \centering
    \includegraphics[width=0.5\textwidth]{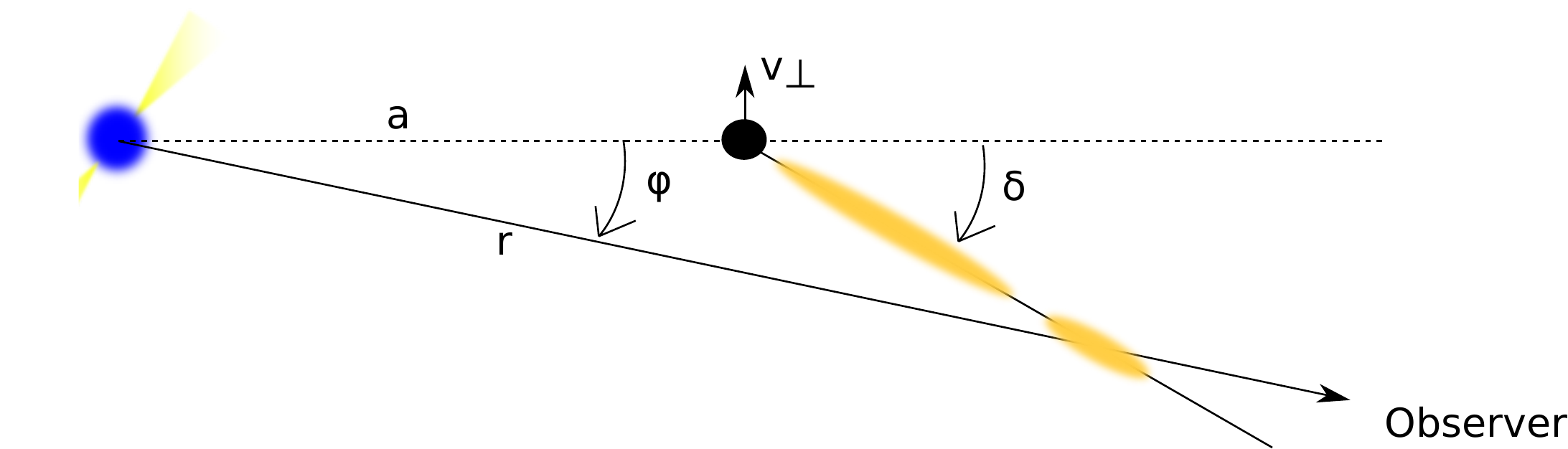}
    \caption{Sketch of an asteroid (black disc) orbiting a pulsar at a distance $a$ (direction of motion according to arrow), and one of its Alv\'en wings trailing with an angle $\delta$. At a given time, an observer sees the part of the wing which is at an angle $\varphi$ from the asteroid position, thus sweeping across emission zones along the wing (orange ellipses). The velocity of the asteroid with respect to the wind in the orthoradial direction, $v_\perp$, fluctuates because of turbulence, being responsible for a wandering a the beam. Turbulence is also responsible for discontinuous emissions zones (yellow ellipses), responsible for sub-bursts.  } %
    \label{fig:wingsweep}
\end{figure}

We consider the Alfv\'en wing structure, as depicted in Fig. \ref{fig:wingsweep}, as the average locus of the emission regions, represented by a straight line at an angle $\delta$ relative to the radial direction at the location of the asteroid. The centre of the observable region is itself at an angle $\varphi = \dot \varphi t$ from the asteroid (since emissions are beamed in the radial direction by the relativistic motion of the wind). Thus as the wing sweeps through the line of sight, the observer sees a region at a distance $r$,
\begin{equation}
\label{eq:rt}
    r \simeq a\left(1+\frac{\varphi}{\tan\delta}\right),
\end{equation}
where we approximated $\varphi \ll 1$. 
Using $\delta \sim \gamma^{-1}$ \citep[][]{mottez_radio_2014}, we get that during a burst the line of sight sweeps 
\begin{equation}
\label{eq:dra}
    \frac{\Delta r}{a} \sim \gamma \dot\varphi \tau = 0.5 \left(\frac{\gamma}{10^6}\right) \left(\frac{\dot \varphi}{10^{-4}}\right)\left(\frac{\tau}{5\rm ms}\right).
\end{equation}
This corresponds to an apparent superluminal speed of $\sim 15000 c$, meaning that the observer receives an instantaneous picture of the wing, with its emissions peaks and lows, as turbulence shapes them. 

We also note that micro-structures have been observed spanning $\sim 10\rm\mu s$ \citep{nimmo_highly_2020}. This could be explained in this framework by smaller clumps of size $\Delta r/a \sim 10^{-3}$, where we used Eq. \eqref{eq:dra} with $\tau = 10 \, \rm\mu s$.

The exact emission mechanism is not specified, but we might assume that it scales with some power $\alpha$ of the magnetic field, as is empirically implied in the radius-to-frequency mapping involved in pulsar emissions, radio maser or synchrotron emissions. Then, we obtain a downward drift in emission frequency $\omega \propto B^\alpha \propto r^{-\alpha}$ such that the variation relative to the emission frequency at the radius of the asteroid is 
\begin{equation}
    \frac{\Delta \omega}{\omega} = \alpha \frac{\Delta r}{a} + \bigcirc\left(\frac{\Delta r}{a}\right)
\end{equation}
at leading order in $r/a$. One sees that the estimate of $\Delta r/a$ obtained in Eq. \eqref{eq:dra} from burst duration is consistent with a linear drift of a few tenth of the highest frequency, as observed for \frb{}, for any moderate value of $\alpha$. This is analogous to the radius-to-frequency mapping invoked in pulsar magnetospheric models \citep[e.g.][]{wang_time-frequency_2019, lyutikov_radius--frequency_2020}.

\subsection{Polarisation}
If we speculate that polarisation angle depends on the local magnetic field, as in pulsar emissions for example, then we do not expect any significant variation of the polarisation angle since the magnetic field in the wind is almost purely azimuthal. Since the radio waves propagate radially, the magnetic field is always perpendicular and does not generate Faraday rotation.

\subsection{Phase shift of the activity window in the LOFAR band}
In the present model, the observed phase shift of $\sim 3$ days in the LOFAR  120 - 180 MHz frequency band compared to the CHIME/FRB band cannot be simply explained by radius-to-frequency mapping using the simple assumptions of Sec. \ref{sec:dfd}. Indeed, assuming $\omega \propto B \propto r^{-1}$ (that is $\alpha=1$), one see that emission in the LOFAR band occurs for $r_{150}/r_{600} \sim 4$, where $r_{\rm xxx}$ is the radius of emission at $\rm xxx$ MHz. Using Eq. \eqref{eq:rt} with $\delta \sim \gamma^{-1}$, one obtains the corresponding time interval $\Delta t \sim 30 \mathrm{ms} \ll 3$ days. Varying the power-law index $\alpha$ does not lead to any more satisfactory result. Rather, this shows that, in the present model, there is an emission cut-off radius for $\Delta r/a \lesssim 1$. 

One may speculate that at large distance from the asteroid, the angle of the Alfv\'en wing $\delta$ (see Fig. \ref{fig:wingsweep}) progressively increases such that the wing curves substantially backwards. This would happen if, for example, the wind were to decelerate while keeping a relatively constant magnetisation. However this runs counter the idea that, as the wind propagates outwards, magnetic reconnection occurs and converts magnetic energy into kinetic energy. 

Another speculation associates the low frequency emissions to an additional circumbinary population of asteroids which, being farther, would generate lower-frequency emissions. The required orbital distance would be compatible with the stability analysis of Sec. \ref{sec:orbstab} assuming radius-to-frequency mapping as above. Emission would be totally independent from higher-frequency bursts, since it would be associated with different asteroids, which would explain the lack of simultaneity of bursts in the two bands as reported in \citep[][]{pleunis_lofar_2021}. However, one hardly sees why the activity window would be shifted, and would rather expect a circumbinary asteroid belt, leading to bursts across all phases and therefore not periodic. Thus, future detection of low-frequency bursts across all phases would strongly support that hypothesis. In order to explain the 10 times larger fluence of the LOFAR bursts \citep[][]{pleunis_lofar_2021}, this would also imply that the circumbinary population be made of much larger asteroids.

\section{Conclusion}
In this article, we have shown that the reported periodicity of \frb as well as properties in terms of peak flux and width can be explained by a relatively low number density of asteroids, a few thousands, immersed in the turbulent and magnetised wind of a young pulsar. We suggest that asteroids are distributed between three dynamical Hilda-type swarms and possibly, but not necessarily, two Trojan swarms, all driven by a main companion. Stability criteria for this orbital configuration lead to an estimate of the main companion in the range $7\times 10^{-4} \lesssim m_c \lesssim 4\times 10^{-3} M_\odot$, compatible with a sub-stellar object such as a brown dwarf.
Although the emission process is not yet specified we have also shown that, assuming some power-law scaling of the emission frequency with the local wind magnetic field, our model can explain downward drifting subpulses, the so-called sad trombone effect \citep[][]{hessels_frb_2019}, as the result of turbulence and radius-to-frequency mapping along the Alfv\'en wing. Micro-structures are similarly explained by smaller scale turbulence, and the observed narrow bandwidth $\Delta \omega / \omega \lesssim 0.5$ is naturally explained by the spread of the emission region in radius as constrained by the duration of the burst. If polarisation is set by the local magnetic field and linear, then the polarisation angle is expected to remain flat across bursts, and without significant rotation measure, implying that the latter is extrinsic to the FRB itself.

 Moreover, our model presents the advantage of providing a number of falsifiable predictions which could be assessed within the next few years if observations remain as frequent as they have been in \citep{collaboration_periodic_2020, pleunis_lofar_2021}:
\begin{itemize}
	\item The main periodicity will become $3P \simeq 49$ days, corresponding to the orbital period of the main companion; 
	\item One of the three groups of FRBs (when folded at $3P$) may become significantly smaller than the others;
	\item Rare bursts might be observed in-between the three favoured orbital phases as Hilda asteroids can be at any orbital phase (only less likely outside of the Lagrange point regions).
	\item Periodic repeaters should become silent after a few decades.
\end{itemize}

Our study favours very young pulsars, at most a millennium old. We note that this is consistent with the apparent location of the source of \frb in a star-forming region of its host galaxy \citep{marcote_repeating_2020}. 

The case of the other periodic repeater FRB121102 is not straightforwardly explained by the current work, partly because the burst distribution with the orbital phase depends on the dynamical history of the system which might very well be different, partly because it is still not completely clear what the period of FRB121102 is due to its longer cycle. Current evidence \citep[][]{rajwade_possible_2020, cruces_repeating_2021} points to a $\sim 160$ day period with a $\sim 50\%$ duty activity cycle. If that is confirmed, then the implied large orbital separation compared to FRB180916, as well as large activity window might be difficult to reconcile with asteroid swarms of the type studied in this work. However, individual burst properties as described in Sec. \ref{sec:individualbursts} apply to repeaters in general.

Finally, we cannot explain in the present work the large phase shift of the activity window in the LOFAR band \citep[][]{pleunis_lofar_2021}. However, should future observations demonstrate bursting activity in the LOFAR band across all phases and not within a small activity window, then this would be a clear signature of an additional circumbinary asteroid belt.

\section*{Acknowledgements}
The authors would like to thank the anonymous reviewer for their insightful and thorough review which helped improving the paper. Many thanks to Dr Melaine Saillenfest for helpful discussions on section 2.4.
G. Voisin acknowledges support of the European Research Council, under the European Union’s Horizon 2020 research and innovation programme (grant agreement No. 715051; Spiders). 
We acknowledge use of the CHIME/FRB Public Database, provided at https://www.chime-frb.ca/ by the CHIME/FRB Collaboration.

\section*{Data Availability}

 This work relies upon the data gathered in the Extended data table 1 of \citet{collaboration_periodic_2020} and \citet{pleunis_lofar_2021}.



\bibliographystyle{mnras}
\bibliography{FRBHildas.bib} 

\begin{thebibliography}{}
\makeatletter
\relax
\def\mn@urlcharsother{\let\do\@makeother \do\$\do\&\do\#\do\^\do\_\do\%\do\~}
\def\mn@doi{\begingroup\mn@urlcharsother \@ifnextchar [ {\mn@doi@}
  {\mn@doi@[]}}
\def\mn@doi@[#1]#2{\def\@tempa{#1}\ifx\@tempa\@empty \href
  {http://dx.doi.org/#2} {doi:#2}\else \href {http://dx.doi.org/#2} {#1}\fi
  \endgroup}
\def\mn@eprint#1#2{\mn@eprint@#1:#2::\@nil}
\def\mn@eprint@arXiv#1{\href {http://arxiv.org/abs/#1} {{\tt arXiv:#1}}}
\def\mn@eprint@dblp#1{\href {http://dblp.uni-trier.de/rec/bibtex/#1.xml}
  {dblp:#1}}
\def\mn@eprint@#1:#2:#3:#4\@nil{\def\@tempa {#1}\def\@tempb {#2}\def\@tempc
  {#3}\ifx \@tempc \@empty \let \@tempc \@tempb \let \@tempb \@tempa \fi \ifx
  \@tempb \@empty \def\@tempb {arXiv}\fi \@ifundefined
  {mn@eprint@\@tempb}{\@tempb:\@tempc}{\expandafter \expandafter \csname
  mn@eprint@\@tempb\endcsname \expandafter{\@tempc}}}

\bibitem[\protect\citeauthoryear{Andersen et~al.,}{Andersen
  et~al.}{2020}]{andersen_bright_2020}
Andersen B.~C.,  et~al., 2020, \mn@doi [Nature] {10.1038/s41586-020-2863-y},
  587, 54

\bibitem[\protect\citeauthoryear{Aschwanden et~al.,}{Aschwanden
  et~al.}{2016}]{aschwanden_25_2016}
Aschwanden M.~J.,  et~al., 2016, \mn@doi [Space Science Reviews]
  {10.1007/s11214-014-0054-6}, 198, 47

\bibitem[\protect\citeauthoryear{Bagchi}{Bagchi}{2017}]{bagchi_unified_2017}
Bagchi M.,  2017, \mn@doi [{\textbackslash}apjl] {10.3847/2041-8213/aa65c9},
  838, L16

\bibitem[\protect\citeauthoryear{Beloborodov}{Beloborodov}{2017}]{beloborodov_flaring_2017}
Beloborodov A.~M.,  2017, \mn@doi [{\textbackslash}apjl]
  {10.3847/2041-8213/aa78f3}, 843, L26

\bibitem[\protect\citeauthoryear{Beloborodov}{Beloborodov}{2020}]{beloborodov_blast_2020}
Beloborodov A.~M.,  2020, \mn@doi [{\textbackslash}apj]
  {10.3847/1538-4357/ab83eb}, 896, 142

\bibitem[\protect\citeauthoryear{Beniamini, Wadiasingh  \& Metzger}{Beniamini
  et~al.}{2020}]{beniamini_periodicity_2020}
Beniamini P.,  Wadiasingh Z.,   Metzger B.~D.,  2020, \mn@doi
  [{\textbackslash}mnras] {10.1093/mnras/staa1783}, 496, 3390

\bibitem[\protect\citeauthoryear{Beutler}{Beutler}{2004}]{beutler_methods_2004}
Beutler G.,  2004, Methods of {Celestial} {Mechanics}: {Volume} {I}:
  {Physical}, {Mathematical}, and {Numerical} {Principles}, 1st softcover
  edition without cd-rom of original hardcover edition. edition edn.
Springer, Berlin ; New York

\bibitem[\protect\citeauthoryear{Bochenek, Ravi  \& Dong}{Bochenek
  et~al.}{2021}]{bochenek_localized_2021}
Bochenek C.~D.,  Ravi V.,   Dong D.,  2021, \mn@doi [The Astrophysical Journal
  Letters] {10.3847/2041-8213/abd634}, 907, L31

\bibitem[\protect\citeauthoryear{Brook, Karastergiou, Buchner, Roberts, Keith,
  Johnston  \& Shannon}{Brook et~al.}{2014}]{brook_evidence_2014}
Brook P.~R.,  Karastergiou A.,  Buchner S.,  Roberts S.~J.,  Keith M.~J.,
  Johnston S.,   Shannon R.~M.,  2014, \mn@doi [The Astrophysical Journal
  Letters] {10.1088/2041-8205/780/2/L31}, 780, L31

\bibitem[\protect\citeauthoryear{Bro{\v z} \& Vokrouhlick{\'y}}{Bro{\v z} \&
  Vokrouhlick{\'y}}{2008}]{broz_asteroid_2008}
Bro{\v z} M.,  Vokrouhlick{\'y} D.,  2008, \mn@doi [Monthly Notices of the
  Royal Astronomical Society] {10.1111/j.1365-2966.2008.13764.x}, 390, 715

\bibitem[\protect\citeauthoryear{Cerutti \& Philippov}{Cerutti \&
  Philippov}{2017}]{cerutti_dissipation_2017}
Cerutti B.,  Philippov A.~A.,  2017, \mn@doi [Astronomy and Astrophysics]
  {10.1051/0004-6361/201731680}, 607, A134

\bibitem[\protect\citeauthoryear{{Chawla} et~al.,}{{Chawla}
  et~al.}{2020}]{2020ApJ...896L..41C}
{Chawla} P.,  et~al., 2020, \mn@doi [\apjl] {10.3847/2041-8213/ab96bf}, \href
  {https://ui.adsabs.harvard.edu/abs/2020ApJ...896L..41C} {896, L41}

\bibitem[\protect\citeauthoryear{Cho et~al.,}{Cho
  et~al.}{2020}]{cho_spectropolarimetric_2020}
Cho H.,  et~al., 2020, \mn@doi [The Astrophysical Journal Letters]
  {10.3847/2041-8213/ab7824}, 891, L38

\bibitem[\protect\citeauthoryear{Collaboration et~al.,}{Collaboration
  et~al.}{2020}]{collaboration_periodic_2020}
Collaboration T.~C.,  et~al., 2020, arXiv:2001.10275 [astro-ph]

\bibitem[\protect\citeauthoryear{Cordes \& Chatterjee}{Cordes \&
  Chatterjee}{2019}]{cordes_fast_2019}
Cordes J.~M.,  Chatterjee S.,  2019, \mn@doi [Annual Review of Astronomy and
  Astrophysics] {10.1146/annurev-astro-091918-104501}, 57, 417

\bibitem[\protect\citeauthoryear{Cordes \& Lazio}{Cordes \&
  Lazio}{2002}]{cordes_ne2001i_2002}
Cordes J.~M.,  Lazio T. J.~W.,  2002, arXiv Astrophysics e-prints, pp
  arXiv:astro--ph/0207156

\bibitem[\protect\citeauthoryear{Cordes \& Shannon}{Cordes \&
  Shannon}{2008}]{cordes_rocking_2008}
Cordes J.~M.,  Shannon R.~M.,  2008, \mn@doi [The Astrophysical Journal]
  {10.1086/589425}, 682, 1152

\bibitem[\protect\citeauthoryear{Cordes \& Wasserman}{Cordes \&
  Wasserman}{2016}]{cordes_supergiant_2016}
Cordes J.~M.,  Wasserman I.,  2016, \mn@doi [{\textbackslash}mnras]
  {10.1093/mnras/stv2948}, 457, 232

\bibitem[\protect\citeauthoryear{Cruces et~al.,}{Cruces
  et~al.}{2021}]{cruces_repeating_2021}
Cruces M.,  et~al., 2021, \mn@doi [Monthly Notices of the Royal Astronomical
  Society] {10.1093/mnras/staa3223}, 500, 448

\bibitem[\protect\citeauthoryear{Dai}{Dai}{2020}]{dai_magnetar-asteroid_2020}
Dai Z.~G.,  2020, \mn@doi [{\textbackslash}apjl] {10.3847/2041-8213/aba11b},
  897, L40

\bibitem[\protect\citeauthoryear{Dai \& Zhong}{Dai \&
  Zhong}{2020}]{dai_periodic_2020}
Dai Z.~G.,  Zhong S.~Q.,  2020, \mn@doi [{\textbackslash}apjl]
  {10.3847/2041-8213/ab8f2d}, 895, L1

\bibitem[\protect\citeauthoryear{Dai, Wang, Wu  \& Huang}{Dai
  et~al.}{2016}]{dai_repeating_2016}
Dai Z.~G.,  Wang J.~S.,  Wu X.~F.,   Huang Y.~F.,  2016, \mn@doi
  [{\textbackslash}apj] {10.3847/0004-637X/829/1/27}, 829, 27

\bibitem[\protect\citeauthoryear{Decoene, Kotera  \& Silk}{Decoene
  et~al.}{2020}]{decoene_fast_2020}
Decoene V.,  Kotera K.,   Silk J.,  2020, arXiv e-prints, 2012,
  arXiv:2012.00029

\bibitem[\protect\citeauthoryear{{Deutsch}}{{Deutsch}}{1955}]{Deutsch_1955}
{Deutsch} A.~J.,  1955, Annales d'Astrophysique, \href
  {http://cdsads.u-strasbg.fr/abs/1955AnAp...18....1D} {18, 1}

\bibitem[\protect\citeauthoryear{Farah et~al.,}{Farah
  et~al.}{2018}]{farah_frb_2018}
Farah W.,  et~al., 2018, arXiv:1803.05697 [astro-ph]

\bibitem[\protect\citeauthoryear{Fonseca et~al.,}{Fonseca
  et~al.}{2020}]{fonseca_nine_2020}
Fonseca E.,  et~al., 2020, \mn@doi [The Astrophysical Journal]
  {10.3847/2041-8213/ab7208}, 891, L6

\bibitem[\protect\citeauthoryear{Gu, Dong, Liu, Ma  \& Wang}{Gu
  et~al.}{2016}]{gu_neutron_2016}
Gu W.-M.,  Dong Y.-Z.,  Liu T.,  Ma R.,   Wang J.,  2016, \mn@doi
  [{\textbackslash}apjl] {10.3847/2041-8205/823/2/L28}, 823, L28

\bibitem[\protect\citeauthoryear{Gu, Yi  \& Liu}{Gu
  et~al.}{2020}]{gu_neutron_2020}
Gu W.-M.,  Yi T.,   Liu T.,  2020, \mn@doi [{\textbackslash}mnras]
  {10.1093/mnras/staa1914}, 497, 1543

\bibitem[\protect\citeauthoryear{Guillemot et~al.,}{Guillemot
  et~al.}{2016}]{guillemot_gamma-ray_2016}
Guillemot L.,  et~al., 2016, \mn@doi [Astronomy \& Astrophysics]
  {10.1051/0004-6361/201527847}, 587, A109

\bibitem[\protect\citeauthoryear{Hessels et~al.,}{Hessels
  et~al.}{2019}]{hessels_frb_2019}
Hessels J. W.~T.,  et~al., 2019, \mn@doi [The Astrophysical Journal]
  {10.3847/2041-8213/ab13ae}, 876, L23

\bibitem[\protect\citeauthoryear{Jewitt, Trujillo  \& Luu}{Jewitt
  et~al.}{2000}]{jewitt_population_2000}
Jewitt D.~C.,  Trujillo C.~A.,   Luu J.~X.,  2000, \mn@doi [The Astronomical
  Journal] {10.1086/301453}, 120, 1140

\bibitem[\protect\citeauthoryear{Kerr, Johnston, Hobbs  \& Shannon}{Kerr
  et~al.}{2015}]{kerr_limits_2015}
Kerr M.,  Johnston S.,  Hobbs G.,   Shannon R.~M.,  2015, \mn@doi [The
  Astrophysical Journal Letters] {10.1088/2041-8205/809/1/L11}, 809, L11

\bibitem[\protect\citeauthoryear{Kotera, Mottez, Voisin  \& Heyvaerts}{Kotera
  et~al.}{2016}]{kotera_asteroids_2016}
Kotera K.,  Mottez F.,  Voisin G.,   Heyvaerts J.,  2016, \mn@doi [Astronomy \&
  Astrophysics] {10.1051/0004-6361/201628116}, 592, A52

\bibitem[\protect\citeauthoryear{Kumar, Lu  \& Bhattacharya}{Kumar
  et~al.}{2017}]{kumar_fast_2017}
Kumar P.,  Lu W.,   Bhattacharya M.,  2017, \mn@doi [{\textbackslash}mnras]
  {10.1093/mnras/stx665}, 468, 2726

\bibitem[\protect\citeauthoryear{Kumar et~al.,}{Kumar
  et~al.}{2020}]{kumar_extremely_2020}
Kumar P.,  et~al., 2020, \mn@doi [Monthly Notices of the Royal Astronomical
  Society] {10.1093/mnras/staa3436}, 500, 2525

\bibitem[\protect\citeauthoryear{Levin, Beloborodov  \& Bransgrove}{Levin
  et~al.}{2020}]{levin_precessing_2020}
Levin Y.,  Beloborodov A.~M.,   Bransgrove A.,  2020, \mn@doi
  [{\textbackslash}apjl] {10.3847/2041-8213/ab8c4c}, 895, L30

\bibitem[\protect\citeauthoryear{Levison, Shoemaker  \& Shoemaker}{Levison
  et~al.}{1997}]{levison_dynamical_1997}
Levison H.~F.,  Shoemaker E.~M.,   Shoemaker C.~S.,  1997, \mn@doi [Nature]
  {10.1038/385042a0}, 385, 42

\bibitem[\protect\citeauthoryear{Lin, Woosley  \& Bodenheimer}{Lin
  et~al.}{1991}]{lin_formation_1991}
Lin D. N.~C.,  Woosley S.~E.,   Bodenheimer P.~H.,  1991, \mn@doi [Nature]
  {10.1038/353827a0}, 353, 827

\bibitem[\protect\citeauthoryear{Lin et~al.,}{Lin et~al.}{2020}]{lin_no_2020}
Lin L.,  et~al., 2020, \mn@doi [Nature] {10.1038/s41586-020-2839-y}, 587, 63

\bibitem[\protect\citeauthoryear{Lu, Kumar  \& Zhang}{Lu
  et~al.}{2020}]{lu_unified_2020}
Lu W.,  Kumar P.,   Zhang B.,  2020, \mn@doi [{\textbackslash}mnras]
  {10.1093/mnras/staa2450}, 498, 1397

\bibitem[\protect\citeauthoryear{Lyne \& Graham-Smith}{Lyne \&
  Graham-Smith}{2012}]{lyne_pulsar_2012}
Lyne A.,  Graham-Smith F.,  2012, Pulsar {Astronomy}, 4 edition edn.
Cambridge University Press, Cambridge ; New York

\bibitem[\protect\citeauthoryear{Lyubarsky}{Lyubarsky}{2014}]{lyubarsky_model_2014}
Lyubarsky Y.,  2014, \mn@doi [Monthly Notices of the Royal Astronomical
  Society] {10.1093/mnrasl/slu046}, 442, L9

\bibitem[\protect\citeauthoryear{Lyutikov}{Lyutikov}{2020}]{lyutikov_radius--frequency_2020}
Lyutikov M.,  2020, \mn@doi [{\textbackslash}apj] {10.3847/1538-4357/ab55de},
  889, 135

\bibitem[\protect\citeauthoryear{Lyutikov, Barkov  \& Giannios}{Lyutikov
  et~al.}{2020}]{lyutikov_frb_2020}
Lyutikov M.,  Barkov M.~V.,   Giannios D.,  2020, \mn@doi [The Astrophysical
  Journal] {10.3847/2041-8213/ab87a4}, 893, L39

\bibitem[\protect\citeauthoryear{Marcote et~al.,}{Marcote
  et~al.}{2020}]{marcote_repeating_2020}
Marcote B.,  et~al., 2020, \mn@doi [Nature] {10.1038/s41586-019-1866-z}, 577,
  190

\bibitem[\protect\citeauthoryear{Marthi, Gautam, Li, Lin, Main, Naidu, Pen  \&
  Wharton}{Marthi et~al.}{2020}]{marthi_detection_2020}
Marthi V.~R.,  Gautam T.,  Li D.~Z.,  Lin H.-H.,  Main R.~A.,  Naidu A.,  Pen
  U.-L.,   Wharton R.~S.,  2020, \mn@doi [Monthly Notices of the Royal
  Astronomical Society] {10.1093/mnrasl/slaa148}, 499, L16

\bibitem[\protect\citeauthoryear{Metzger, Berger  \& Margalit}{Metzger
  et~al.}{2017}]{metzger_millisecond_2017}
Metzger B.~D.,  Berger E.,   Margalit B.,  2017, \mn@doi [The Astrophysical
  Journal] {10.3847/1538-4357/aa633d}, 841, 14

\bibitem[\protect\citeauthoryear{Michtchenko \& Ferraz-Mello}{Michtchenko \&
  Ferraz-Mello}{1996}]{michtchenko_comparative_1996}
Michtchenko T.~A.,  Ferraz-Mello S.,  1996, Astronomy and Astrophysics, 310,
  1021

\bibitem[\protect\citeauthoryear{Moons}{Moons}{1996}]{moons_review_1996}
Moons M.,  1996, \mn@doi [Celestial Mechanics and Dynamical Astronomy]
  {10.1007/BF00048446}, 65, 175

\bibitem[\protect\citeauthoryear{{Mottez} \& {Heyvaerts}}{{Mottez} \&
  {Heyvaerts}}{2011}]{Mottez_2011_AWW}
{Mottez} F.,  {Heyvaerts} J.,  2011, \mn@doi [Astronomy and Astrophysics]
  {10.1051/0004-6361/201116530}, \href
  {http://cdsads.u-strasbg.fr/abs/2011A%26A...532A..21M} {532, A21+}

\bibitem[\protect\citeauthoryear{Mottez \& Zarka}{Mottez \&
  Zarka}{2014}]{mottez_radio_2014}
Mottez F.,  Zarka P.,  2014, \mn@doi [Astronomy and Astrophysics]
  {10.1051/0004-6361/201424104}, 569, A86

\bibitem[\protect\citeauthoryear{Mottez, Zarka  \& Voisin}{Mottez
  et~al.}{2020}]{mottez_repeating_2020}
Mottez F.,  Zarka P.,   Voisin G.,  2020, \mn@doi [Astronomy \& Astrophysics]
  {10.1051/0004-6361/202037751}

\bibitem[\protect\citeauthoryear{Murray \& Dermott}{Murray \&
  Dermott}{1999}]{murray_solar_1999}
Murray C.~D.,  Dermott S.~F.,  1999, Solar system dynamics.
Cambridge, Royaume-Uni de Grande-Bretagne et d'Irlande du Nord, Etats-Unis
  d'Am{\'e}rique

\bibitem[\protect\citeauthoryear{Nakamura \& Piran}{Nakamura \&
  Piran}{1991}]{nakamura_origin_1991}
Nakamura T.,  Piran T.,  1991, \mn@doi [The Astrophysical Journal Letters]
  {10.1086/186217}, 382, L81

\bibitem[\protect\citeauthoryear{{Neubauer}}{{Neubauer}}{1980}]{neubauer_1980}
{Neubauer} F.~M.,  1980, \mn@doi [\jgr] {10.1029/JA085iA03p01171}, \href
  {https://ui-adsabs-harvard-edu.insu.bib.cnrs.fr/abs/1980JGR....85.1171N} {85,
  1171}

\bibitem[\protect\citeauthoryear{Nimmo et~al.,}{Nimmo
  et~al.}{2020}]{nimmo_highly_2020}
Nimmo K.,  et~al., 2020, arXiv e-prints, 2010, arXiv:2010.05800

\bibitem[\protect\citeauthoryear{Pastor-Marazuela et~al.,}{Pastor-Marazuela
  et~al.}{2020}]{pastor-marazuela_chromatic_2020}
Pastor-Marazuela I.,  et~al., 2020, arXiv e-prints, 2012, arXiv:2012.08348

\bibitem[\protect\citeauthoryear{Petit, Laskar  \& Bou{\'e}}{Petit
  et~al.}{2017}]{petit_amd-stability_2017}
Petit A.~C.,  Laskar J.,   Bou{\'e} G.,  2017, \mn@doi [Astronomy and
  Astrophysics] {10.1051/0004-6361/201731196}, 607, A35

\bibitem[\protect\citeauthoryear{P{\'e}tri}{P{\'e}tri}{2016}]{petri_theory_2016}
P{\'e}tri J.,  2016, \mn@doi [Journal of Plasma Physics]
  {10.1017/S0022377816000763}, 82, 635820502

\bibitem[\protect\citeauthoryear{Petroff, Hessels  \& Lorimer}{Petroff
  et~al.}{2019}]{petroff_fast_2019}
Petroff E.,  Hessels J. W.~T.,   Lorimer D.~R.,  2019, \mn@doi [The Astronomy
  and Astrophysics Review] {10.1007/s00159-019-0116-6}, 27, 4

\bibitem[\protect\citeauthoryear{{Phinney} \& {Hansen}}{{Phinney} \&
  {Hansen}}{1993}]{phinney_pulsar_1993}
{Phinney} E.~S.,  {Hansen} B.~M.~S.,  1993, in {Phillips} J.~A.,  {Thorsett}
  S.~E.,   {Kulkarni} S.~R.,  eds,  Astronomical Society of the Pacific
  Conference Series Vol. 36, Planets Around Pulsars. pp 371--390

\bibitem[\protect\citeauthoryear{Pleunis et~al.,}{Pleunis
  et~al.}{2021}]{pleunis_lofar_2021}
Pleunis Z.,  et~al., 2021, \mn@doi [The Astrophysical Journal Letters]
  {10.3847/2041-8213/abec72}, 911, L3

\bibitem[\protect\citeauthoryear{{Podsiadlowski}}{{Podsiadlowski}}{1993}]{podsiadlowski_planet_1993}
{Podsiadlowski} P.,  1993, in {Phillips} J.~A.,  {Thorsett} S.~E.,   {Kulkarni}
  S.~R.,  eds,  Astronomical Society of the Pacific Conference Series Vol. 36,
  Planets Around Pulsars. pp 149--165

\bibitem[\protect\citeauthoryear{Popov \& Postnov}{Popov \&
  Postnov}{2010}]{popov_hyperflares_2010}
Popov S.~B.,  Postnov K.~A.,  2010, in Harutyunian H.~A.,  Mickaelian A.~M.,
  Terzian Y.,  eds, Evolution of {Cosmic} {Objects} through their {Physical}
  {Activity}. pp 129--132

\bibitem[\protect\citeauthoryear{Popov \& Postnov}{Popov \&
  Postnov}{2013}]{popov_millisecond_2013}
Popov S.~B.,  Postnov K.~A.,  2013, arXiv e-prints, 1307, arXiv:1307.4924

\bibitem[\protect\citeauthoryear{Rajwade et~al.,}{Rajwade
  et~al.}{2020}]{rajwade_possible_2020}
Rajwade K.~M.,  et~al., 2020, arXiv:2003.03596 [astro-ph]

\bibitem[\protect\citeauthoryear{Shannon et~al.,}{Shannon
  et~al.}{2013}]{shannon_asteroid_2013}
Shannon R.~M.,  et~al., 2013, \mn@doi [The Astrophysical Journal]
  {10.1088/0004-637X/766/1/5}, 766, 5

\bibitem[\protect\citeauthoryear{{Sicardy} \& {Lissauer}}{{Sicardy} \&
  {Lissauer}}{1992}]{Sicardy_1992}
{Sicardy} B.,  {Lissauer} J.~J.,  1992, \mn@doi [Advances in Space Research]
  {10.1016/0273-1177(92)90425-W}, \href
  {https://ui.adsabs.harvard.edu/abs/1992AdSpR..12...81S} {12, 81}

\bibitem[\protect\citeauthoryear{Smallwood, Martin  \& Zhang}{Smallwood
  et~al.}{2019}]{smallwood_investigation_2019}
Smallwood J.~L.,  Martin R.~G.,   Zhang B.,  2019, \mn@doi
  [{\textbackslash}mnras] {10.1093/mnras/stz483}, 485, 1367

\bibitem[\protect\citeauthoryear{{Spitler} et~al.,}{{Spitler}
  et~al.}{2018}]{spitler_2018}
{Spitler} L.~G.,  et~al., 2018, \mn@doi [\apj] {10.3847/1538-4357/aad332},
  \href
  {https://ui-adsabs-harvard-edu.insu.bib.cnrs.fr/abs/2018ApJ...863..150S}
  {863, 150}

\bibitem[\protect\citeauthoryear{Tendulkar et~al.,}{Tendulkar
  et~al.}{2021}]{tendulkar_60_2021}
Tendulkar S.~P.,  et~al., 2021, \mn@doi [The Astrophysical Journal Letters]
  {10.3847/2041-8213/abdb38}, 908, L12

\bibitem[\protect\citeauthoryear{Wadiasingh \& Chirenti}{Wadiasingh \&
  Chirenti}{2020}]{wadiasingh_fast_2020}
Wadiasingh Z.,  Chirenti C.,  2020, \mn@doi [{\textbackslash}apjl]
  {10.3847/2041-8213/abc562}, 903, L38

\bibitem[\protect\citeauthoryear{Wadiasingh \& Timokhin}{Wadiasingh \&
  Timokhin}{2019}]{wadiasingh_repeating_2019}
Wadiasingh Z.,  Timokhin A.,  2019, \mn@doi [{\textbackslash}apj]
  {10.3847/1538-4357/ab2240}, 879, 4

\bibitem[\protect\citeauthoryear{Wang, Chakrabarty  \& Kaplan}{Wang
  et~al.}{2006}]{wang_debris_2006}
Wang Z.,  Chakrabarty D.,   Kaplan D.~L.,  2006, \mn@doi [Nature]
  {10.1038/nature04669}, 440, 772

\bibitem[\protect\citeauthoryear{Wang, Zhang, Chen  \& Xu}{Wang
  et~al.}{2019}]{wang_time-frequency_2019}
Wang W.,  Zhang B.,  Chen X.,   Xu R.,  2019, \mn@doi [{\textbackslash}apjl]
  {10.3847/2041-8213/ab1aab}, 876, L15

\bibitem[\protect\citeauthoryear{Wolszczan}{Wolszczan}{2012}]{wolszczan_discovery_2012}
Wolszczan A.,  2012, \mn@doi [New Astronomy Reviews]
  {10.1016/j.newar.2011.06.002}, 56, 2

\bibitem[\protect\citeauthoryear{Wolszczan \& Frail}{Wolszczan \&
  Frail}{1992}]{wolszczan_planetary_1992}
Wolszczan A.,  Frail D.~A.,  1992, \mn@doi [Nature] {10.1038/355145a0}, 355,
  145

\bibitem[\protect\citeauthoryear{Yuan, Beloborodov, Chen  \& Levin}{Yuan
  et~al.}{2020}]{yuan_plasmoid_2020}
Yuan Y.,  Beloborodov A.~M.,  Chen A.~Y.,   Levin Y.,  2020, \mn@doi [The
  Astrophysical Journal] {10.3847/2041-8213/abafa8}, 900, L21

\bibitem[\protect\citeauthoryear{Zanazzi \& Lai}{Zanazzi \&
  Lai}{2020}]{zanazzi_periodic_2020}
Zanazzi J.~J.,  Lai D.,  2020, \mn@doi [{\textbackslash}apjl]
  {10.3847/2041-8213/ab7cdd}, 892, L15

\bibitem[\protect\citeauthoryear{Zarka}{Zarka}{2020}]{Zarka2020}
Zarka P.,  2020, Star-Planet Interactions in the Radio Domain: Prospect for
  Their Detection.
Springer International Publishing, Cham, pp 1--16,
  \mn@doi{10.1007/978-3-319-30648-3_22-2}, \url
  {https://doi.org/10.1007/978-3-319-30648-3_22-2}

\bibitem[\protect\citeauthoryear{collaboration}{collaboration}{2013}]{collaboration_second_2013}
collaboration T. F.-L.,  2013, \mn@doi [The Astrophysical Journal Supplement
  Series] {10.1088/0067-0049/208/2/17}, 208, 17

\bibitem[\protect\citeauthoryear{{de Pater} et~al.,}{{de Pater}
  et~al.}{2005}]{dePater_2005}
{de Pater} I.,  et~al., 2005, \mn@doi [\icarus] {10.1016/j.icarus.2004.10.020},
  \href {https://ui.adsabs.harvard.edu/abs/2005Icar..174..263D} {174, 263}

\makeatother
\end{thebibliography}




\appendix

\section{Pulsar and companions characteristics compatible with \frb} \label{etude_parametrique}
We conducted a parametric study in order to check if the MZV20 model fits the characteristics of \frb when the companion orbital periods are set to $P, 2P$ and $3P$.
The model and notations are exactly the same as in MZV20: a pulsar of $1.4$ Solar mass, with a surface magnetic field $B_*$, a surface temperature $T_*$, a radius $R_*$ and a rotation period $P_*$, emits a power $\dot E_{\mathrm{max}}$ (also noted $ (1-f) g\dot E_\mathrm{rp}$ in MZV20) in the form of high energy photons and wind kinetic energy, that contributes to the companion thermal balance. The power $\dot E_{\mathrm{max}}$ is less than the loss rate of rotational energy, because an important part of the rotational loss is in the form of the long-wavelength Poynting flux \citep{Deutsch_1955} that is not absorbed by small companions \citep{kotera_asteroids_2016}.  The companion orbits at a distance $a$ ($r$ in MZV20) from the neutron star, where it is immersed in the pulsar  wind of Lorentz factor $\gamma$. The companion is heated by the thermal radiation of the pulsar associated with $T_*$, by the high-energy photons and particles of the wind (both associated with $\dot E_{\mathrm{max}}$), and by the electric current induced by the Alfv\'en wings into the companion. This last contribution depends on its electrical conductivity $\sigma_c$, where $\sigma_c \sim 10^3$ for silicate rocks, and $\sigma_c \sim 10^7$ for an iron dominated body.

The radio emission power is proportional to the electromagnetic power associated with the Alfv\'en wing \citep{Mottez_2011_AWW}  with a yield coefficient $\epsilon \sim 10^{-3} - 10^{-2}$ (\citet{mottez_radio_2014}, MZV20, \citet{Zarka2020} and references therein). 
In the reference frame of the radio source, that is the pulsar wind reference frame, we supposed that the waves are emitted within a solid angle $\Omega_A$. In the observer's frame, this solid angle is considerably reduced by the relativistic aberration. Regarding radio frequencies, the emission bandwidth is thought to be 1 GHz. The FRB duration $\tau$ is used in the computation of the source size (in our reference frame).

We considered a distance $D = 0.15$ Gpc between the source and the observer \citep{marcote_repeating_2020}. Since the bursts detected by CHIME/FRB \citep{collaboration_periodic_2020,pleunis_lofar_2021} have  a flux $0.3< S<6.6$ Jy, we retained parameter sets corresponding to pulses exceeding 0.3 Jy from the distance $D=0.15$ Gpc.

The procedure is the same as in MZV20 : we tried the 1336500 combinations of parameters displayed in Table \ref{table_parametrique_pulsar3}, that correspond to bodies orbiting pulsars with a high magnetic field. We then selected the cases that meet the following conditions: 
(1) the observed signal amplitude on Earth must exceed 0.3 Jy; (2) the companion must be in solid state with no melting/evaporation happening; (3) the radius of the source must exceed the maximum local Larmor radius. This last condition is a condition of validity of the MHD equations that support the theory of Alfv\'en wings.{ Practically, a smaller Larmor radius might be associated with electron and positrons. In our analysis, this radius} is compiled for hydrogen ions at the speed of light, so condition (3) is checked conservatively. 

Among those parameter sets, 140382 fit our conditions. Because we were interested in small companions and by fairly energetic pulsars, we then required  $R_c \le 10$ km, and  $\dot E_{\mathrm{max}} \ge 10^{27}$ W, and a pulsar spin-down age $\tau=P_*/\dot P_*> 10$ years (see MZV20 for the evaluation of $\tau$ in our analysis). We then got 197   solutions for an orbital period $P_{\mathrm{orb}}=P$,  299   for Hildas companions ($P_{\mathrm{orb}}=2P$) and 252 for Trojans ($P_{\mathrm{orb}}=3P$). All of them exclude a period $P_* \geq 30$ ms, only $P_*=10$ ms and $P_*=3$ ms are retained. A few of them are detailed in Table \ref{table_jeux_de_parametres_Rc_petit}.

By definition, all the solutions correspond to a flux above 0.3 Jy, and 20 of the 299 solutions  associated with Hildas companions correspond to bursts above 10 Jy.
All of them involve metal rich companions, with $\sigma_C \ge 100   $  Mho. We found solutions down to $R_c \le 2 $ km. The neutron star temperature does not constrain very much the solutions, and $T_* = 3 \times 10^6   $ can be reached without problem.  The smallest obtained wind Lorentz factor is $\gamma = 3 \times 10^5   $. The smallest magnetic field $B_*= 3.2 \times 10^7$ T. The non-thermal radiations are more constraining :  the largest found value is $\dot E_{\mathrm{max}} \ge 10^{28}$ W. The spin-down age in years is comprised in the range $11.4       \le \tau \le 642.3     $. Thus, only young pulsars can cause Hildas or Trojans FRBs associated with \frb.

There are 252    solutions  associated with Trojan companions with orbital period $3P$, and 12 of them correspond to bursts above 10 Jy.  Their characteristics are very similar to those of Hildas with orbital period $2P$ ; they exhibit almost the same maximum and minimum values. 

The thermal constraints are computed for circular orbits. For Hildas, with a period $2P$ a corresponding semi-major axis $a=0.22$ AU, and eccentricity $e \sim 0.3$ (in the Solar system $0\leq e \leq 0.3$), the periastron distance is $r_p=0.15$ is similar to the semi-major axis $a=0.14$ corresponding to a circular orbit of period $P$. Therefore, the thermal constraints of Hildas are somewhere between the cases $P$ and $2P$ of Table \ref{table_jeux_de_parametres_Rc_petit}, and the constraints relative to the radio-emission power are those corresponding to a period $2P$.

In any case, we could see that FRBs caused by small companions ($R_c$ down to 2 km) are associated with a highly magnetized pulsar with periods about 3 or 10 ms. This class of pulsars is represented in our Galaxy by the Crab pulsar and by younger pulsars.

\begin{table*}
	\centering 
\begin{tabular}{|c|l|l|l|} \hline
 Input parameters	& Notation & Values 	& Unit		\\ \hline
NS magnetic field 		& $B_*$		& $10^7,   3.2 \times 10^7,   10^8,   3.2 \times 10^8,   10^9   $ 		&	T	 			\\
NS radius				& $R_*$		&  $10,   11,   12,   13   $ 		&	km	 				\\
Rotation period			& $P_*$		& $3.2 \times 10^{-3},   10^{-2},   3.2 \times 10^{-2},   10^{-1},   3.2 \times 10^{-1}   $ 	& s	 					\\
NS temperature			& $T_*$		& $3. \times 10^5,   10^6,   3.0 \times 10^6   $		&	 K			\\
Wind Lorentz factor		& $\gamma$	&$10^5   3\times 10^5   10^6   $	&		 				\\
Radio 	efficiency		& $\epsilon$& $ 10^{-2}   $& 	 	  				\\
Companion orbital period	& $P_{\mathrm{orb}}$		& $P, 2P, 3P$	&	$P=16.34$ d	 				\\
Companion radius		& $R_c$		& $10,   22,  46,  100   $	&	m	 		\\
Companion radius		& $R_c$		& $3.2 \times 10^2,   10^3,   3.2 \times 10^3   10^4   $	&	m	 		\\
Emission solid angle 	& $\Omega_A$		& $0.1,   1,   10   $ &	sr	 		\\
Power input			& $\dot E_{\mathrm{max}}$& $10^{27},   3 \times 10^{27},   10^{28},   3 \times 10^{28},   10^{29}   $ 		&	W	 		\\
Companion conductivity	& $\sigma_c$		& $10^{-3},   10^2   10^7   $ 		&Mho	 		\\
Distance to observer	& $D$		&$0.15$ 		&	Gpc	 			\\
Bandwidth				& $\Delta f$& 1 		&	GHz	 			\\
FRB duration			& $\tau$		& $5. \, 10^{-3}$ 		&	s	 			\\
\hline
\end{tabular}
	\caption{Parameter set of the first parametric study of FRBs produced by pulsar companions of medium and small size.} 
	\label{table_parametrique_pulsar3} 
\end{table*}

\begin{table*}
	\begin{tabular}{|c|l|l|l|l|l|l|l|l|l|l|l|l|l|l|l|l|} 
		\hline 
Parameter   &  $B_*$     & $R_*$   &  $P_*$ &   $\gamma$ &    $a$     &   $R_c$   & $\dot E_{\mathrm{max}}$& $\sigma_C$ & $S$ & $\tau$\\
Unit        &  T         & km      & s      &            &   AU       &   km      & W         & Mho    & Jy & yr\\
\hline
long $\tau$ &$ 3 \times 10^7  $&$ 10  $&$ 0.003 $&$ 1 \times 10^6 $&$ 0.141   $&$ 10     $&$ 3 \times 10^{27}  $&$ 10^7  $&$ 0.5      $&$ 1137.9   $\\ 
large $S$ &$ 3 \times 10^8  $&$ 10  $&$ 0.003 $&$ 1 \times 10^6 $&$ 0.141   $&$ 10     $&$ 3 \times 10^{27}  $&$ 10^7  $&$ 53.6     $&$ 11.4     $\\ 
long $P_*$ &$ 3 \times 10^8  $&$ 12  $&$ 0.010 $&$ 1 \times 10^6 $&$ 0.141   $&$ 10     $&$ 3 \times 10^{27}  $&$ 10^7  $&$ 1.6      $&$ 38.1     $\\ 
\hline
low $\gamma$ &$ 1 \times 10^8  $&$ 11  $&$ 0.003 $&$ 3 \times 10^5 $&$ 0.224   $&$ 10     $&$ 3 \times 10^{27}  $&$ 10^7  $&$ 0.3      $&$ 64.2     $\\ 
large $\tau$  &$ 3 \times 10^7  $&$ 11  $&$ 0.003 $&$ 1 \times 10^6 $&$ 0.224   $&$ 10     $&$ 1 \times 10^{28}  $&$ 10^7  $&$ 0.4      $&$ 642.3    $\\ 
large $S$ &$ 3 \times 10^8  $&$ 10  $&$ 0.003 $&$ 1 \times 10^6 $&$ 0.224   $&$ 10     $&$ 1 \times 10^{28}  $&$ 10^2  $&$ 21.3     $&$ 11.4     $\\ 
small $R_c$  &$ 3 \times 10^8  $&$ 10  $&$ 0.003 $&$ 1 \times 10^6 $&$ 0.224   $&$ 2      $&$ 1 \times 10^{27}  $&$ 10^7  $&$ 1.0      $&$ 11.4     $\\ 
longer $P_*$ &$ 1 \times 10^9  $&$ 10  $&$ 0.010 $&$ 1 \times 10^6 $&$ 0.224   $&$ 10     $&$ 1 \times 10^{27}  $&$ 10^2  $&$ 2.1      $&$ 11.4     $\\ 
longer $P_*$ &$ 3 \times 10^8  $&$ 11  $&$ 0.010 $&$ 1 \times 10^6 $&$ 0.224   $&$ 10     $&$ 1 \times 10^{27}  $&$ 10^2  $&$ 0.4      $&$ 64.2     $\\ 
\hline
low $\gamma$  &$ 1 \times 10^8  $&$ 12  $&$ 0.003 $&$ 3 \times 10^5 $&$ 0.293   $&$ 10     $&$ 1 \times 10^{28}  $&$ 10^2  $&$ 0.3      $&$ 38.1     $\\ 
low $\gamma$ &$ 3 \times 10^8  $&$ 10  $&$ 0.003 $&$ 3 \times 10^5 $&$ 0.293   $&$ 10     $&$ 1 \times 10^{27}  $&$ 10^7  $&$ 1.1      $&$ 11.4     $\\ 
large $\tau$ &$ 3 \times 10^7  $&$ 12  $&$ 0.003 $&$ 1 \times 10^6 $&$ 0.293   $&$ 10     $&$ 1 \times 10^{28}  $&$ 10^2  $&$ 0.4      $&$ 381.1    $\\ 
large $S$ &$ 3 \times 10^8  $&$ 10  $&$ 0.003 $&$ 1 \times 10^6 $&$ 0.293   $&$ 10     $&$ 1 \times 10^{28}  $&$ 10^2  $&$ 12.4     $&$ 11.4     $\\ 
small $R_c$ &$ 3 \times 10^8  $&$ 10  $&$ 0.003 $&$ 1 \times 10^6 $&$ 0.293   $&$ 2      $&$ 1 \times 10^{27}  $&$ 10^7  $&$ 0.6      $&$ 11.4     $\\ 
\hline
	\end{tabular}
	\caption{Examples illustrating the results of the parametric studies in Tables \ref{table_parametrique_pulsar3} for pulsars with small companions.  In all cases, the NS temperature is $T_*=10^6$ K.} 
	\label{table_jeux_de_parametres_Rc_petit} 
	\centering 
\end{table*}


\bsp	
\label{lastpage}
\end{document}